\documentclass{iopart}
\usepackage[T1]{fontenc}
\usepackage[latin9]{inputenc}
\usepackage{graphicx}
\usepackage{esint}
\usepackage[titletoc]{appendix}

\makeatletter

\usepackage{iopams}
\usepackage{setstack}

\makeatother

\begin{document}

\title{Mixed order transition and condensation in exactly soluble one dimensional
spin model}

\author{Amir Bar, David Mukamel}
\address{Department of Physics of Complex Systems, Weizmann Institute of Science, Rehovot 7610001, Israel}
\ead{amir.bracha@weizmann.ac.il, david.mukamel@weizmann.ac.il}
\begin{abstract}
Mixed order phase transitions (MOT), which display discontinuous
order parameter and diverging correlation length, appear
in several seemingly unrelated settings ranging from equilibrium
models with long-range interactions to models far from thermal equilibrium.
In a recent paper \cite{bar2014mixed} an exactly soluble spin model with long-range
interactions that exhibits MOT was introduced and analyzed both by
a grand canonical calculation and a renormalization group analysis.
The model was shown to lay a bridge between two classes of one dimensional
models exhibiting MOT, namely between spin models with inverse distance square
interactions and surface depinning models. In this paper we elaborate
on the calculations done in \cite{bar2014mixed}. We also analyze
the model in the canonical ensemble, which yields a better insight into
the mechanism of MOT. In addition, we generalize the model to include
Potts and general Ising spins, and also consider a broader class of
interactions which decay with distance with a power law different from 2.
\end{abstract}
\maketitle
\tableofcontents{}

\section{Introduction}

Classification of phase transitions is a narrative in statistical
physics and quantum field theory. A common starting point is the distinction
between first order transitions and continuous - or critical - transitions.
While in first order transitions the order parameter is discontinuous,
or more generally the free energy is non-differentiable,
in continuous transitions the order parameter changes continuously
and the free energy is non-analytic but has a first derivative. Continuous
transitions are known to possess universal features and can thus be categorized
into universality classes, as opposed to first order transitions
which are non-universal. The universality of critical transitions is tightly related
to the divergence of a correlation length. Much progress has
been made in classifying critical transitions based on methods such
as renormalization group and conformal field theory which rely on
such divergence.

However, there are known examples that deviate from the above scheme,
for instance transitions for which on the one hand the free energy
is non-differentiable but on the other hand they display a diverging correlation
length \cite{fisher1982scaling}. Early examples of such mixed order
transitions (MOT) include (a) one dimensional discrete spin models
with long-range interactions \cite{anderson1969exact,thouless1969long,dyson1971ising,cardy1981one,aizenman1988discontinuity,slurink1983roughening},
and (b) models of depinning transitions such as models of DNA denaturation
\cite{PS1966,fisher1966effect} and of wetting \cite{blossey1995diverging,fisher1984walks}.
More recently there has been renewed interest in MOT taking place
in studies of percolation models in the context of glass and jamming transitions
\cite{gross1985mean,toninelli2006jamming,toninelli2007toninelli,schwarz2006onset,liu2012core},
evolution of complex networks \cite{liu2012extraordinary,zia2012extraordinary,tian2012nature,bizhani2012discontinuous}
and active biopolymer gels \cite{sheinman2014discontinuous}. While
all these examples exhibit MOT, they display some qualitative and quantitative
differences. For instance the divergence of the correlation length
is algebraic in some of them, while it is a stretched exponential in
others. It would be interesting to understand the origin of the similarities
and differences between the various models by studying them within a common framework.

In \cite{bar2014mixed} a new model which establishes a link between
the classes of models (a) and (b) was introduced
and analyzed, thus making a step towards a unified view. Specifically,
this study focuses on two specific representative models of classes
(a) and (b): The one dimensional Ising model with interactions decaying
with the distance $r$ between spins as $r^{-2}$ (dubbed IDSI, for
inverse distance squared Ising model) \cite{thouless1969long,anderson1970exact,aizenman1988discontinuity}
and the Poland Scheraga (PS) model for DNA denaturation \cite{PS1966,KMP2000}.
The first class exhibits a Kosterlitz-Thouless (KT) vortex unbinding
like transition \cite{kosterlitz1973ordering}, while the second class
exhibits a condensation transition similar to the Bose Einstein condensation
(BEC) transition of free bosons. The model presented in \cite{bar2014mixed},
which we shall refer to as the Truncated IDSI (TIDSI) model, thus
provides an intriguing connection between KT and BEC transitions in
one dimension. Another appealing property of the TIDSI model is that
it is exactly soluble.

In this paper we extend the analysis of \cite{bar2014mixed} in several
directions:
\begin{itemize}
\item The calculations done in \cite{bar2014mixed} are elaborated, and
many features of the phase diagram which where only briefly mentioned
in \cite{bar2014mixed} are derived explicitly.
\item The exact free energies of different ensembles are derived. These
shed new light on the origin of criticality in the mixed-order transition appearing in
the TIDSI model.
\item We generalize the model to include spins other than Ising spins, so
that these models could be compared with models of class (a) other than the
IDSI \cite{cardy1981one}. We also consider long-range interactions
decaying as $r^{-\alpha}$, with $\alpha\neq2$.
\end{itemize}
The paper is organized as follows: we start in section
\ref{sec:Mixed-order-transitions} by reviewing MOT in one dimension and describe
in detail the IDSI and PS models. Then in section \ref{sec:The-model}
the TIDSI model is presented, and the relation to the IDSI and PS
model is discussed. We also summarize in this section the main results
of the paper, including the phase diagram and the free energies. The
next three sections are dedicated to deriving these results using
grand canonical analysis (section \ref{sec:Grand-canonical-analysis}),
canonical analysis (section \ref{sec:Canonical-analysis}) and
renormalization group (RG) analysis (section \ref{sec:RG-analysis}).
While the TIDSI model is exactly soluble, the IDSI model is not, but as
we show in section \ref{sec:RG-analysis} the RG analysis
can be used as a common framework for studying both models. The TIDSI model is then extended to include more general
interactions (section \ref{sub:General-interactions-decay}) and other
spin variables (section \ref{sub:General-spins}).

\section{\label{sec:Mixed-order-transitions}Mixed order transitions (MOT)
in one dimension}

In this section we briefly review two models exhibiting MOT in one
dimension. As mentioned above these models are prototypical representatives
of wider classes of models.

\subsection{The inverse distance squared Ising (IDSI) model}

\begin{figure}
\centering{}\includegraphics[scale=0.5]{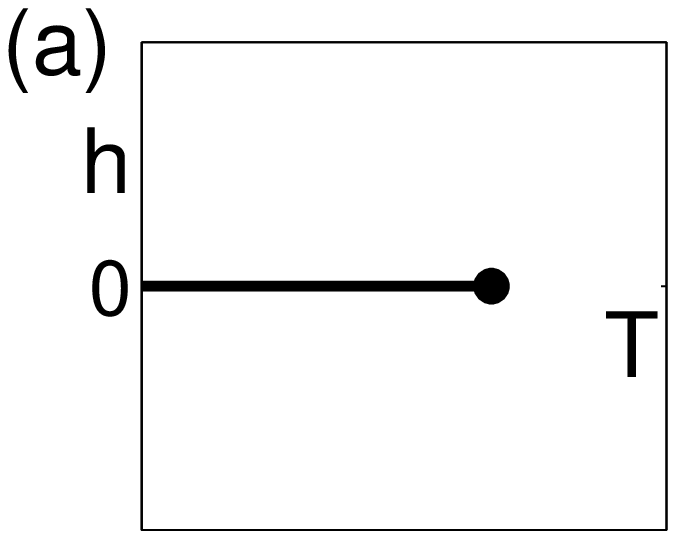}\includegraphics[scale=0.5]{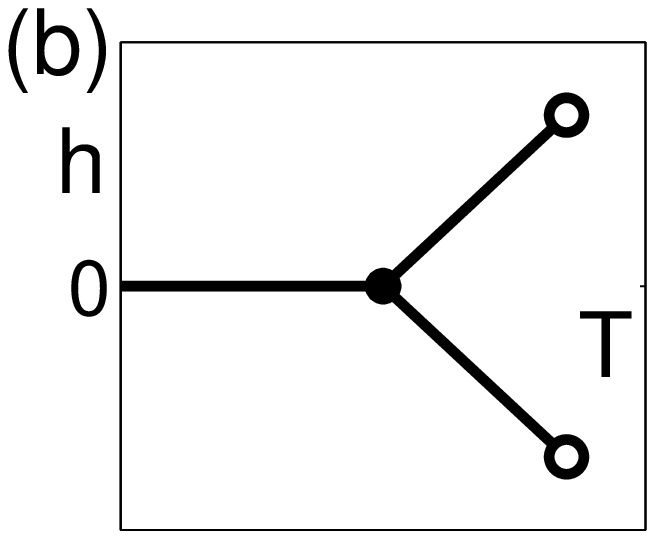}\caption{\label{fig:IDSI_PT} Schematic phase diagrams of (a) the IDSI model
defined by Eq.(\ref{eq_1D:IDSI}), and (b) the Blume-Capel model define
by Eq.(\ref{eq_1D:MFS1I}).}
\end{figure}

The IDSI is a one dimensional Ising model with long-range interactions
which decay asymptotically as $r^{-2}$ where $r$ is the distance
between the spins. Formally it is defined on a one dimensional lattice
of size $L$, where in each lattice site there is a spin variable
$\sigma_{i}=\pm1$, and the Hamiltonian is
\begin{equation}
H=-\sum_{i<j}J(i-j)\sigma_{i}\sigma_{j}\qquad;\qquad J(r\gg1)\sim r^{-2},\label{eq_1D:IDSI}
\end{equation}
with $J(r)\ge0$ for all $r$. More generally, one can consider interactions
decaying as $J(r\gg1)\sim r^{-\alpha}$. It
has been proved by Ruelle \cite{ruelle1968statistical} and Dyson
\cite{dyson1969existence} that $\alpha=2$ is a borderline case,
i.e. that models with $\alpha>2$ exhibit no phase transition, while
models with $1<\alpha<2$ show a symmetry breaking transition at some
finite critical temperature $T_{c}>0$. Models with $\alpha\le1$
have non-extensive energy and hence they are ordered at all temperatures.
We shall henceforth refer to the $\alpha=2$ case as the inverse distance
squared Ising (IDSI) model. Thouless, through an entropy-energy argument,
has suggested that the IDSI model exhibits a unique phase transition
in which the magnetization exhibits a discontinuity at the critical
temperature \cite{thouless1969long}. Dyson proved the existence of
such a transition, (where the discontinuity of the order parameter
is dubbed the ``Thouless effect''), in a hierarchical version of
the model \cite{dyson1971ising}. A scaling analysis carried out by Anderson
et al. \cite{anderson1970exact} and a later Renormalization group
(RG) analysis done by Cardy \cite{cardy1981one}, showed that the
transition in the IDSI model is of the Kosterlitz-Thouless (KT) type
\cite{kosterlitz1973ordering}, with correlation length which diverges
at the transition with an essential singularity as $\xi\sim\exp\left(1/\sqrt{T-T_{c}}\right)$.
%Historically, the scaling analysis of Anderson et al. actually preceded
%and laid the foundations for Kosterlitz and Thouless analysis of the
%XY model.
The rigorous proof for the existence of the transition in
the IDSI model was given only in the 1980's by Frohlich and Spencer
\cite{frohlich1982phase}, while the mixed order nature of the transition
was proved substantially later by Aizenman et al. \cite{aizenman1988discontinuity}.
A numerical validation for the analytical predictions was
given in \cite{luijten1997classical,luijten2001criticality}.

The phase diagram of the IDSI model in the temperature-magnetic field
$\left(T,h\right)$ plane is given in Fig.\ref{fig:IDSI_PT}a. It
resembles the $d>1$ dimensional Ising model phase diagram, with a
low-temperature first order transition at zero magnetic field $h$,
which terminates at a critical point at $T_{c}>0$. The only qualitative
difference is that at $T=T_{c}$ the transition of the IDSI is discontinuous,
that is the order parameter --- the magnetization --- jumps from $0$
at $T=T_{c}^{+}$ to $\pm m_{c}\neq0$ at $T=T_{c}^{-}$. However,
the transition, as mentioned above, is still critical as there is
a diverging correlation length. This is different from more common
first order symmetry breaking transitions, as is found, for instance,
in Blume-Capel model
\begin{equation}
H=-J\sum_{i\neq j}\sigma_{i}\sigma_{j}+\Delta\sum_{i}\sigma_{i}^{2}-h\sum_{i}\sigma_{i}\qquad;\qquad\sigma_{i}\in\left\{ -1,0,1\right\} .\label{eq_1D:MFS1I}
\end{equation}
in a certain range of the parameters $J$ and $\Delta$ and at $h=0$.

A schematic temperature-magnetic field phase diagram for the Blume-Capel
model for fixed $J$ and $\Delta$ is presented in Fig.\ref{fig:IDSI_PT}b.
Here the $T=T_{c}$, $h=0$ point is actually a triple point in which
three first order lines meet, i.e. in which three phases coexist.
These are the $+$ phase, the $-$ phase and the disordered phase.
There are also finite $h$ transition lines on which there is a coexistence
of two magnetically ordered phases. These finite $h$ lines terminate
in a usual second order critical point \cite{fisher1982scaling}.

Although the main features of the phase diagram of the IDSI model have been proven rigorously
\cite{frohlich1982phase,aizenman1988discontinuity}, there are many
missing details: Non-universal quantities
such as the critical temperature are not known exactly. The free energy
of the model was not calculated, and clearly the partition function
cannot be calculated in any ensemble. The model which is analyzed
in this paper is a modified version of the IDSI in which all of these
quantities can be exactly calculated.

\subsection{The Poland-Scheraga (PS) model}

\begin{figure}
\begin{centering}
\includegraphics[scale=0.5]{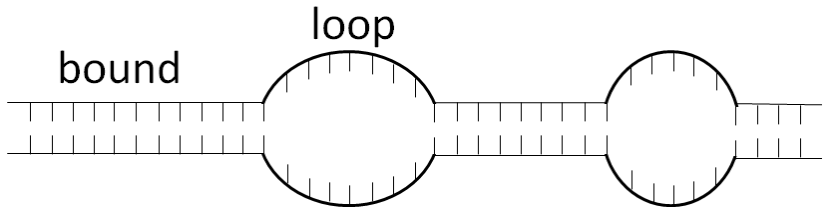}
\par\end{centering}

\caption{\label{fig:PS_config}A typical configuration of the PS model}
\end{figure}

\begin{figure}
\begin{centering}
\includegraphics[scale=0.8]{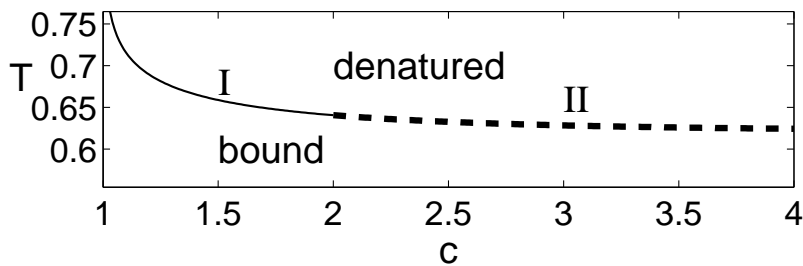}
\par\end{centering}

\caption{\label{fig:PS_PT}Phase diagram of the PS model.}
\end{figure}

The PS model is a model of DNA denaturation \cite{PS1966}. Denaturation
is the process in which the two strands of the double-stranded DNA
molecule separate upon heating. The PS model idealizes the DNA molecule
as an alternating chain of bound segments and denatured loops as depicted
in Fig.\ref{fig:PS_config}. For homopolymers, while a bound segment
of length $l$ contributes an energy $-\epsilon l$ , $\epsilon>0$
being the binding energy of DNA base-pairs, loops are assumed
to contribute no energy, but they contribute entropy $S$ of the form
$e^{S}\approx\omega\frac{s^{l}}{l^{c}}$, where $\omega$ and $s$ are some non-universal constants and $c$
is a universal exponent termed the \emph{loop exponent}. Thus
\begin{equation}
S(l)=bl-c\log\left(l\right)+\tilde{\Delta},\label{eq_1D:PS_ent}
\end{equation}
with $b\equiv\log\left(s\right)$ and $\tilde{\Delta}\equiv\log\left(\omega\right)$.
This form emerges from modeling the denatured loop as a closed path
\cite{fisher1966effect}, which results in a universal exponent $c$ that depends only
on dimensionality and topological constraints, i.e. whether the path
is considered as an ideal walk, self avoiding walk etc. \cite{KMP2000}.
The order parameter for this system is the fraction of bound base-pairs
$\theta$, with $\theta=0$ above the melting temperature $T_{c}$,
and is positive below $T_{c}$. The nature of the transition depends
on the exponent $c$, as depicted in the phase diagram in Fig.\ref{fig:PS_PT}:
For $c\le1$ the strands are bound at all temperatures and
$T_{c}=\infty$. For $1<c\le2$ there
is a continuous transition at $T=T_{c}$ with $\theta\rightarrow0$
as $T\rightarrow T_{c}$. For $c>2$ the order parameter is discontinuous,
i.e. $\theta\rightarrow\theta_{c}>0$ as $T\rightarrow T_{c}$. For
all $c>1$ the correlation length $\xi$ diverges at the transition
point as $\xi\sim\left(T-T_{c}\right)^{-\nu}$ with $\nu=\max\left\{ \frac{1}{c-1},1\right\} $,
and hence for $c>2$ the transition is of MOT type. The mechanism of the transition
is mathematically similar to the Bose-Einstein condensation (BEC)
transition in non-interacting Bose gas, with the condensate being
the macroscopic loop formed at the transition temperature $T_{c}$.

There are several clear differences between the MOT transition in the PS model and in
the IDSI model. First, the correlation length in the two models diverges differently:  algebraically in the PS
model and as a stretched exponential in the IDSI model. Within a renormalization group analysis, the PS model
displays a one-parameter family of fixed points with continuously varying exponent while the transition in the IDSI is
controlled by a single fixed point. In addition, the PS transition is
a condensation transition which is characterized by the formation of a single macroscopic
loop. No such macroscopic object is associated with the IDSI transition.
Finally, while the IDSI transition is a symmetry breaking transition, there
is no symmetry which is broken in the PS case. To understand the origin
of the similarities and differences between these two models --- and
between the classes they represent --- we introduced in \cite{bar2014mixed}
the TIDSI model which serves as a bridge between these models. We
now turn to define the model and show explicitly its similarity to
those two models.

\section{\label{sec:The-model}TIDSI model}

\subsection{Definition}

Like the IDSI model, the TIDSI model of \cite{bar2014mixed} is defined on a one dimensional Ising chain, where on each site $i$
there is a spin variable $\sigma_{i}=\pm1$. The interaction between spins is composed
of a nearest neighbor interaction term $-J_{NN}\sigma_{i}\sigma_{i+1}$
and a long-range interaction term $-J(i-j)\sigma_{i}\sigma_{j}I\left(i\sim j\right)$,
where $I\left(i\sim j\right)=1$ if sites $i$ and $j$ are in the
same domain of either all up or all down spins, and $I\left(i\sim j\right)=0$
otherwise. Thus the long-range interactions (beyond nearest neighbors)
are confined within each of the domains. As in the IDSI model we take the interaction
strength $J(r)$ to decay as $r^{-2}$. The indicator
function $I\left(i\sim j\right)$ can be written explicitly in terms of the spin variables
\[
I\left(i\sim j\right)=\prod_{k=i}^{j-1}\delta_{\sigma_{k}\sigma_{k+1}}=\prod_{k=i}^{j-1}\frac{1+\sigma_{k}\sigma_{k+1}}{2}.
\]
Hence the resulting Hamiltonian reads
\begin{equation}
H=-J_{NN}\sum\sigma_{i}\sigma_{i+1}-\sum_{i<j}J(i-j)\sigma_{i}\sigma_{j}\prod_{k=i}^{j-1}\frac{1+\sigma_{k}\sigma_{k+1}}{2},\label{eq_model:ttt_Hamiltonian1}
\end{equation}
which explicitly contains multi-spin interactions of arbitrary order.
For concreteness we take $J(r)=Cr^{-2}$ for $r\ge1$, even though
the results are qualitatively the same for any $J(r)$ which has the
same asymptotic form.

An alternative representation of a configuration of the model is in
terms of domains: Each microscopic configuration is defined in terms
of the spin at the first site $\sigma_{1}$, the number of domains
$1\le N\le L$ and their lengths $\left\{ l_{a}\right\} _{a=1}^{N}$
where $l_{a}\ge1$ and $\sum_{a}l_{a}=L$. Due to the truncation of
the long-range interactions to within domains, the energy (\ref{eq_model:ttt_Hamiltonian1})
can straightforwardly be expressed in terms of these variables: The nearest
neighbors term sums up to
\begin{equation}
H_{NN}=-J_{NN}\left[\left(L-1\right)-2\left(N-1\right)\right]=2J_{NN}N-J_{NN}\left(L+1\right),\label{eq_model:H_NN}
\end{equation}
while the energy contribution of the long-range term for each domain
$H_{LR}\left(l\right)$ is given by
\begin{eqnarray}
H_{LR}\left(l\right) & = & -C\sum_{k=1}^{l}\frac{l-k}{k^{2}}=-C\left[l\left(\zeta_{2}-\frac{a}{l}+O\left(l^{-2}\right)\right)-\log\left(l\right)+O\left(l^{-1}\right)\right]\nonumber \\
 & = & -bl+C\log\left(l\right)+\tilde{\Delta}+O\left(l^{-1}\right),\label{eq_model:H_LR}
\end{eqnarray}
where $\zeta_{2}\equiv\sum_{k=1}^{\infty}k^{-2}$, $a>0$ is an expansion
coefficient, $b\equiv C\zeta_{2}$ and $\tilde{\Delta}\equiv Ca$.
Eq.(\ref{eq_model:H_LR}) is of the same form as $-TS\left(l\right)$
where $S$ is the loop entropy in the PS model given in Eq.(\ref{eq_1D:PS_ent}),
under the mapping $c\rightarrow\beta C$ where $\beta=T^{-1}$ (we
use units in which the Boltzmann constant $k_{B}$ is unity). In essence,
the physics of both models stem from the logarithmic dependence of
the energy (entropy) on the domain (loop) length. The mapping between
energy and entropy implies an inversion of the temperature role, and
hence of a mapping between the high temperature phase of the PS model
and the low temperature phase of the TIDSI model.
Other than this trivial difference, the phenomenology of the two models
is almost identical. The only qualitative difference between the phase
diagrams of the two models is due to the additional spin inversion
symmetry which exists in the TIDSI model and is lacking in the PS
model, as will be discussed below. Another important difference is
that while the parameter $c$ in the PS model is a universal exponent,
for the TIDSI model the corresponding parameter $\beta_{c}C$, with
$\beta_{c}=T_{c}^{-1}$, depends on the details of
the model such as the nearest-neighbors coupling $J_{NN}$.

\begin{figure}
\begin{centering}
\includegraphics[scale=0.5]{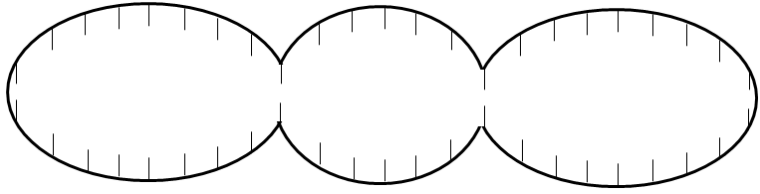}
\par\end{centering}

\caption{\label{fig:ttt_PS_map}An illustration of the TIDSI model as a symmetric
PS model, where all spin are within loops (domains) and none are in
bound segments (compare to Fig.\ref{fig:PS_config}).}
\end{figure}

Neglecting terms of order $O\left(l^{-1}\right)$ and combining (\ref{eq_model:H_NN})
and (\ref{eq_model:H_LR}), the Hamiltonian expressed by the domains
variables reads
\begin{equation}
H\left(N;\left\{ l_{n}\right\} \right)=H_{NN}+\sum_{n}H_{LR}\left(l_{n}\right)=\Delta N+C\sum_{n}\log\left(l_{n}\right)+Const,\label{eq_model:ttt_Hamiltonian2}
\end{equation}
where $\Delta\equiv\tilde{\Delta}+2J_{NN}>0$. The constant term in
(\ref{eq_model:ttt_Hamiltonian2}), which contains also expressions
like $-J_{NN}L$, can be set to $0$ without loss of generality, and
this will be our convention henceforth. In terms of the mapping to
the PS model, the Hamiltonian (\ref{eq_model:ttt_Hamiltonian2}) can
be viewed as the log-Boltzmann factor of a configuration made up of a sequence of loops
without bound segments in between, as depicted in Fig.\ref{fig:ttt_PS_map}.
Hence the TIDSI model can be viewed either as a truncated version
of the IDSI model or as a symmetrized version of the PS model.

The transition in the TIDSI model can be described using two order parameters: first,
the magnetization $m=\frac{1}{L}\sum\sigma_{i}$, which in the domains
variables reads
\begin{equation}
m=-\frac{\sigma_{1}}{L}\sum_{a=1}^{N}\left(-1\right)^{a}l_{a}.\label{eq_model:domains_mag}
\end{equation}
The second order parameter is the density of domains $n\equiv\frac{N}{L}$.
A magnetic field $h$ can be added to the Hamiltonian (\ref{eq_model:ttt_Hamiltonian2})
by adding a term $-hM$ where $M\equiv Lm$, and similarly a loop chemical
potential $\mu$ can be added as $-\mu N$.

Finally, we should define the statistical ensemble which we shall
focus on. Both the IDSI and PS models are defined in the canonical
ensemble in which the length of the chain $L$ is fixed. The partition
function of the TIDSI model in this ensemble can be written in terms
of the domains variables as
\begin{equation}
Z_{C}\left(L;\beta\right)=\sum_{N=1}^{\infty}\sum_{l_{1}=1}^{\infty}...\sum_{l_{N}=1}^{\infty}e^{-\beta H}I\left(L=\sum_{a=1}^{N}l_{a}\right),\label{eq_model:Z_C_def}
\end{equation}
with $H$ given by (\ref{eq_model:ttt_Hamiltonian2}) and the indicator
function $I\left(\psi\right)=1$ if $\psi$ is true and is $0$
otherwise. However, we also consider more restricted ensembles in
which either the magnetization $M$ or number of domains $N$ or both
are fixed. The free energies in such ensembles play a similar role
of a Landau free energy, enabling deeper understanding of the nature
of the transition, as we discuss below. The partition function in
the ensemble in which both $M$ and $N$ are fixed reads
\begin{equation}
Z_{0}\left(L,M,N;\beta\right)=\sum_{\left\{ l_{a}\right\} }e^{-\beta H}I\left(L=\sum l_{a}\right)I\left(M=Lm\left(\left\{ l_{a}\right\} \right)\right).\label{eq_model:Z_0_def}
\end{equation}
From this partition function we can derive the marginal partition
sums
\begin{eqnarray}
Z_{M}\left(L,M;\beta\right) & = & \sum_{N}Z_{0}\left(L,M,N;\beta\right),\label{eq_model:Z_M_def}\\
Z_{N}\left(L,N;\beta\right) & = & \sum_{M}Z_{0}\left(L,M,N;\beta\right),\label{eq_model:Z_N_def}\\
Z_{C}\left(L;\beta\right) & = & \sum_{M,N}Z_{0}\left(L,M,N;\beta\right).
\end{eqnarray}

\subsection{Summary of results}

In this section we present a summary of the results derived for the TIDSI model in sections
\ref{sec:Grand-canonical-analysis} and \ref{sec:Canonical-analysis}.
As stated above, the transition in the TIDSI model is quite similar
to the transition of the PS model (only the role of temperature is
inverted): The high temperature phase of the TIDSI model is composed of a gas
of microscopic (finite size) domains, while the low temperature
phase consists essentially of a single macroscopic domain. The
transition is then a condensation transition, it is MOT for zero magnetic
field but can be either continuous or MOT for nonzero field, as discussed in details in the next subsection.
In addition to calculating exactly the phase diagram, we calculated
the free energies of the model in ensembles in which either the magnetization
or the density of domains are fixed. These free energies play the
role of the Landau free energies in some sense, and hence allow a
deeper understanding of the behavior of the system, as discussed in
subsection \ref{sub:Free-energies}.

\subsubsection{Phase diagram }

\begin{figure}
\includegraphics[scale=0.5]{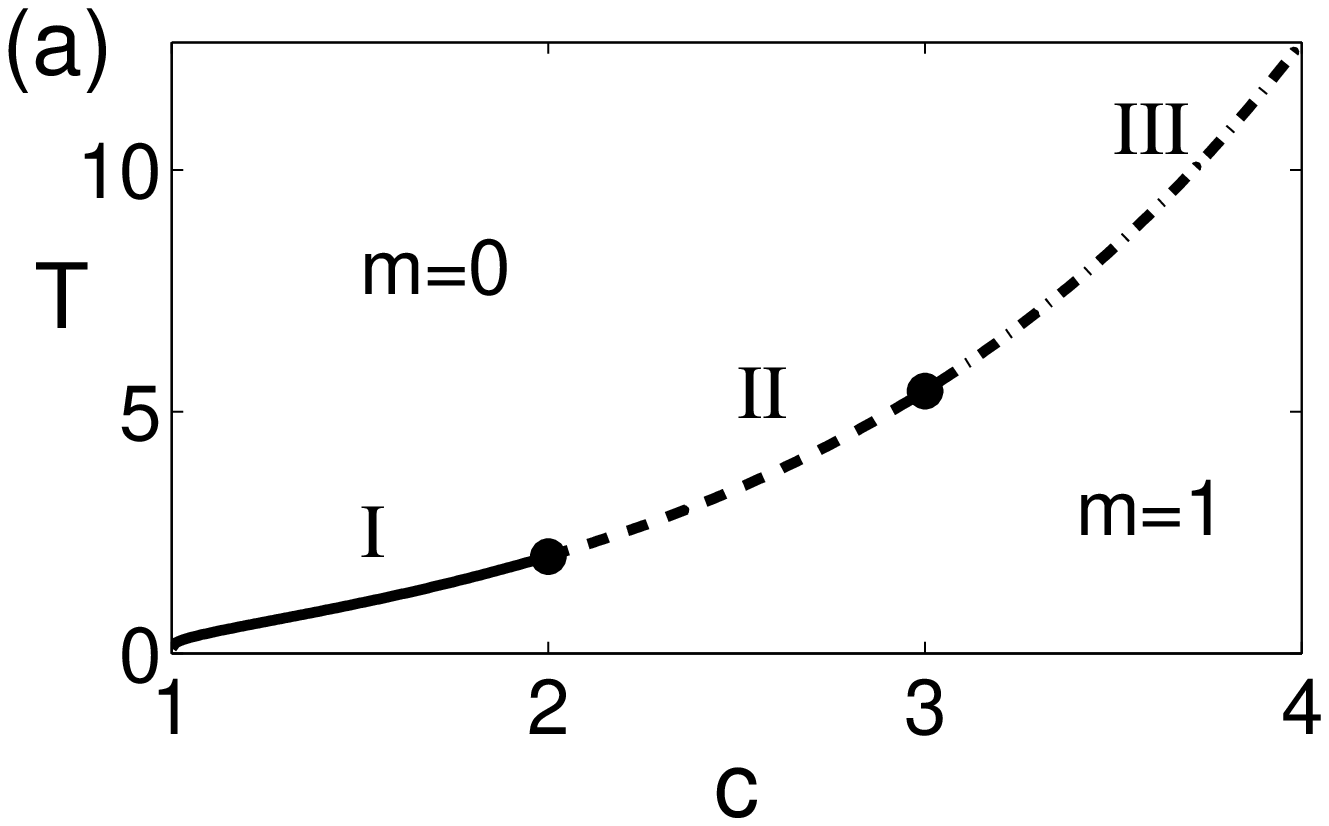}\includegraphics[scale=0.5]{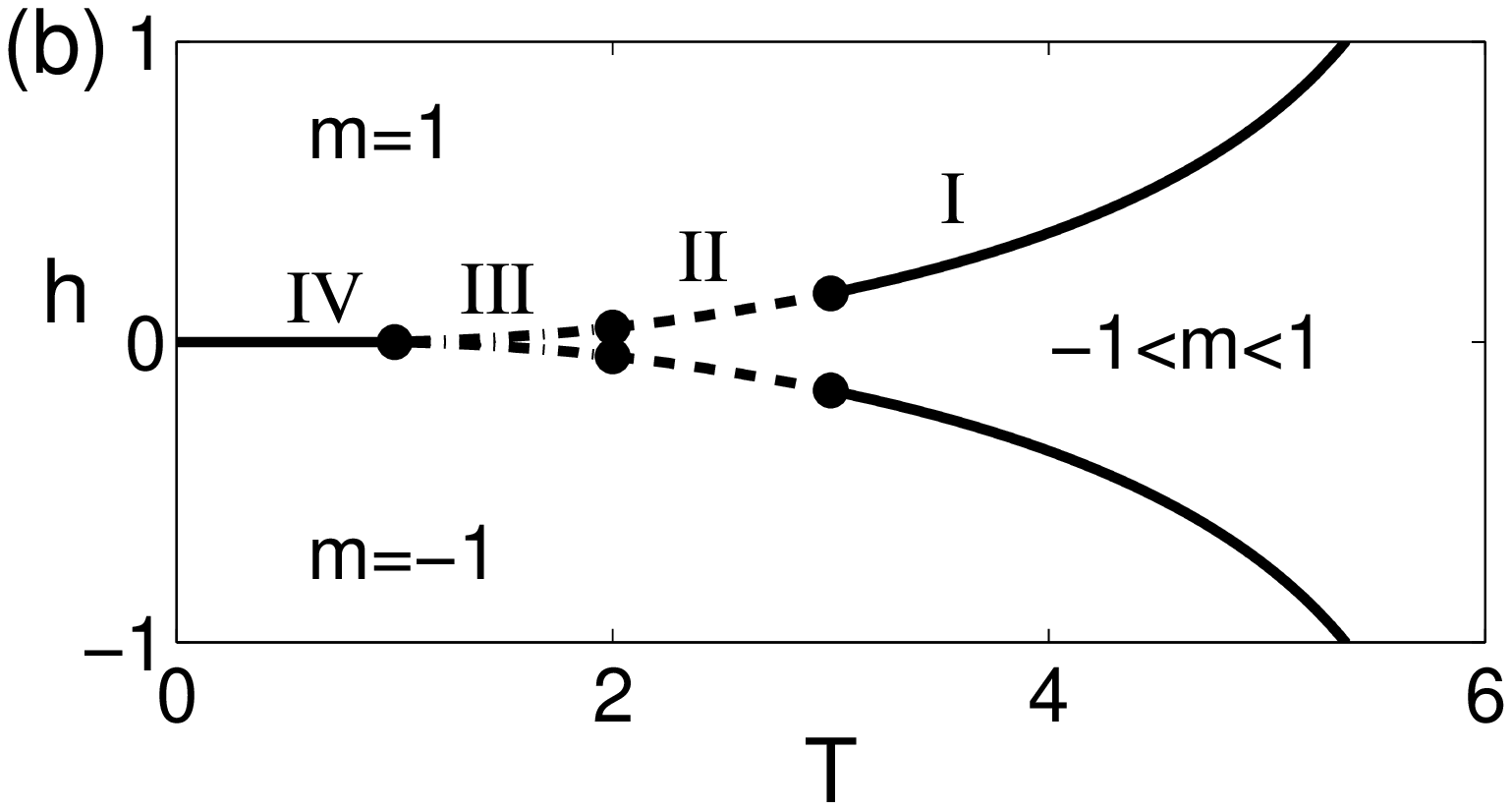}
\caption{\label{fig:Phase_diagram}Phase diagram of the TIDSI model: (a) the $\left(c,T\right)$ projection, with $\Delta=1$; and (b)
the $\left(T,h\right)$ projection for $c=C=6$ and $\Delta=\log\zeta_{c}$
so that $T_{c}\left(h=0\right)=1$. The different regions (I-IV) are
explained in the text.}
\end{figure}

The phase diagram of the TIDSI model is presented in Fig.\ref{fig:Phase_diagram}.
Fig.\ref{fig:Phase_diagram}a displays the transition temperature
($T_{c}$) as a function of the coupling $c$ at $h=0$. Here
\[
c\equiv\beta_{c}C,
\]
where $\beta_{c}=T_{c}^{-1}$ and $T_{c}$ is the critical temperature.
The transition line is given by
\begin{equation}
\zeta\left(\beta_{c}C\right)=e^{\beta_{c}\Delta},\label{eq:results_betac}
\end{equation}
where $\zeta\left(z\right)$ is the Riemann Zeta function. The high
temperature phase is disordered. Hence the average magnetization
vanishes ,$\left\langle m\right\rangle =0$, and the number of domains
is macroscopic $\left\langle n\right\rangle >0$. For $T<T_{c}$,
the phase is ordered, but unlike spin models with two-body interactions,
the magnetization is saturated, i.e. $\left\langle m\right\rangle =\pm1$
throughout the low temperature phase. In addition $\left\langle n\right\rangle =0$
as there is essentially a single macroscopic domain. This jump between
zero to saturated magnetization is an extreme example of the Thouless
effect. As mentioned above, the Thouless effect --- first conjectured to
take place in the IDSI model --- refers to a transition in which
the magnetization is discontinuous while the
correlation length diverges. In considering the other order parameter
$\left\langle n\right\rangle $, one finds that the nature of the
transition changes along the transition line: for $1<c\equiv\beta_{c}C\le2$
(region I) the average density of the domains $\left\langle n\right\rangle $
drops continuously to $0$; for $2<c\le3$ (region II) there is a
discontinuity in $\left\langle n\right\rangle $
from a finite density of domains $n_{c}$ to $0$ at the transition, but the magnetic
susceptibility $\chi_{0}\equiv\left.\frac{\partial^{2}\log Z_{C}}{\partial h^{2}}\right|_{h=0}$
diverges as $T\searrow T_{c}$; for $c>3$ (region III) the density
of domains $\left\langle n\right\rangle $ is discontinuous and $\chi_{0}$
is finite. For any $c>1$ the correlation length $\xi$ diverges as
\begin{equation}
\xi\sim\left(T-T_{c}\right)^{-\nu}\quad;\quad\nu=\max\left(1,\frac{1}{c-1}\right).\label{eq_res:xi}
\end{equation}
The mechanism of the transition is similar to that of the Bose-Einstein
condensation (BEC): while above the critical temperature there is
an extensive number of microscopic domains, which can be referred
to as a normal gas of domains, below $T_{c}$ there is a single macroscopic
domain, which can be referred to as the condensate.

In Fig.\ref{fig:Phase_diagram}b the phase diagram in the $\left(T,h\right)$
plane is plotted. The transition lines in the $\left(T,h\right)$
diagram are given by
\begin{equation}
\zeta\left(\beta_{c}C\right)\Phi_{\beta_{c}C}\left(e^{-2\beta_{c}\left|h\right|}\right)=e^{2\beta_{c}\Delta},
\end{equation}
where $\Phi_{\gamma}(u)$ is the Polylogarithm function, which satisfies
$\Phi_{\gamma}(1)=\zeta\left(\gamma\right)$. Each of the two transition
lines at finite $h$ separates two phases: The gas phase where $-1<\left\langle m\right\rangle <1$
and $\left\langle n\right\rangle >0$, and the condensed phases where
$\left\langle n\right\rangle =0$ and $\left\langle m\right\rangle =1$
or $\left\langle m\right\rangle =-1$. The nature of the transition
changes along the transition line: for $c(h)\equiv\beta_{c}(h)C>3$ (region III) there is a diverging
length scale, finite susceptibility and discontinuous $\left\langle m\right\rangle $ and $\left\langle n\right\rangle $. For $2<c(h)\le3$ (region II) the susceptibility
becomes divergent. For $1<c(h)\le2$ (region I) the susceptibility diverges and the transition is continuous
both in $\left\langle n\right\rangle $ and $\left\langle m\right\rangle $.
This is unlike the $h=0$ case, in which the magnetization is discontinuous.
For $T<T_{c}(h)$ the transition line between the two condensed phases
(region IV) is a normal first order transition as in short-range Ising
models (in $d\ge2$).

\subsubsection{\label{sub:Free-energies}Free energies}

In addition to the phase diagram, the logarithm of the partition function
was evaluated to leading and next to leading order in $L^{-1}$. The
leading order term is just the free energy, or more generally the
large deviations function (LDF) \cite{touchette2009large}. The next
to leading order (finite size correction) provides an insight into
the mechanism of the transition, revealing a logarithmic (in $L$)
barrier between coexisting phases, which implies that the fluctuations of
the order parameters decay as a power-law and hence are critical.

\paragraph{Large Deviations functions}

\begin{figure}
\includegraphics[scale=0.5]{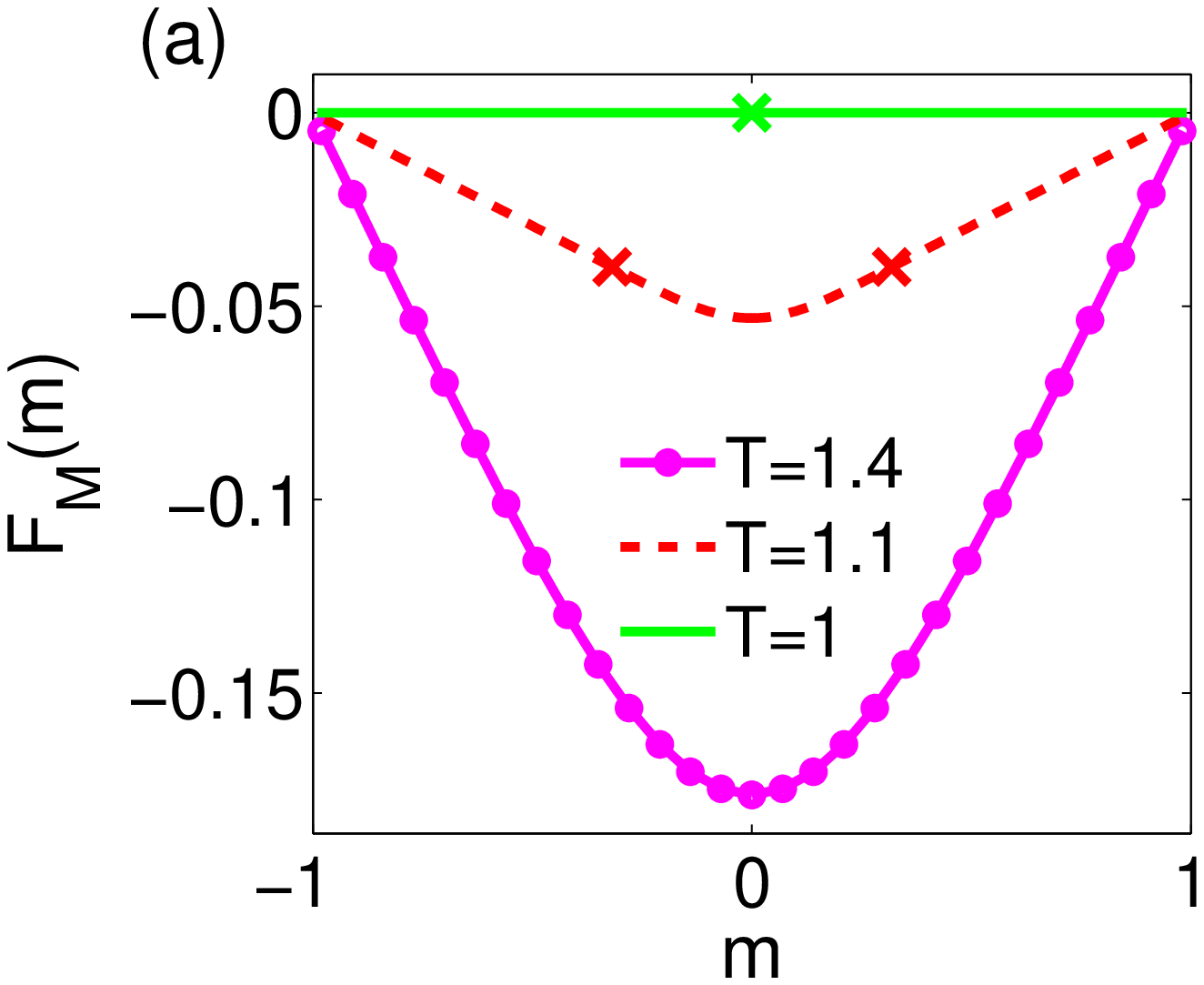}\includegraphics[scale=0.5]{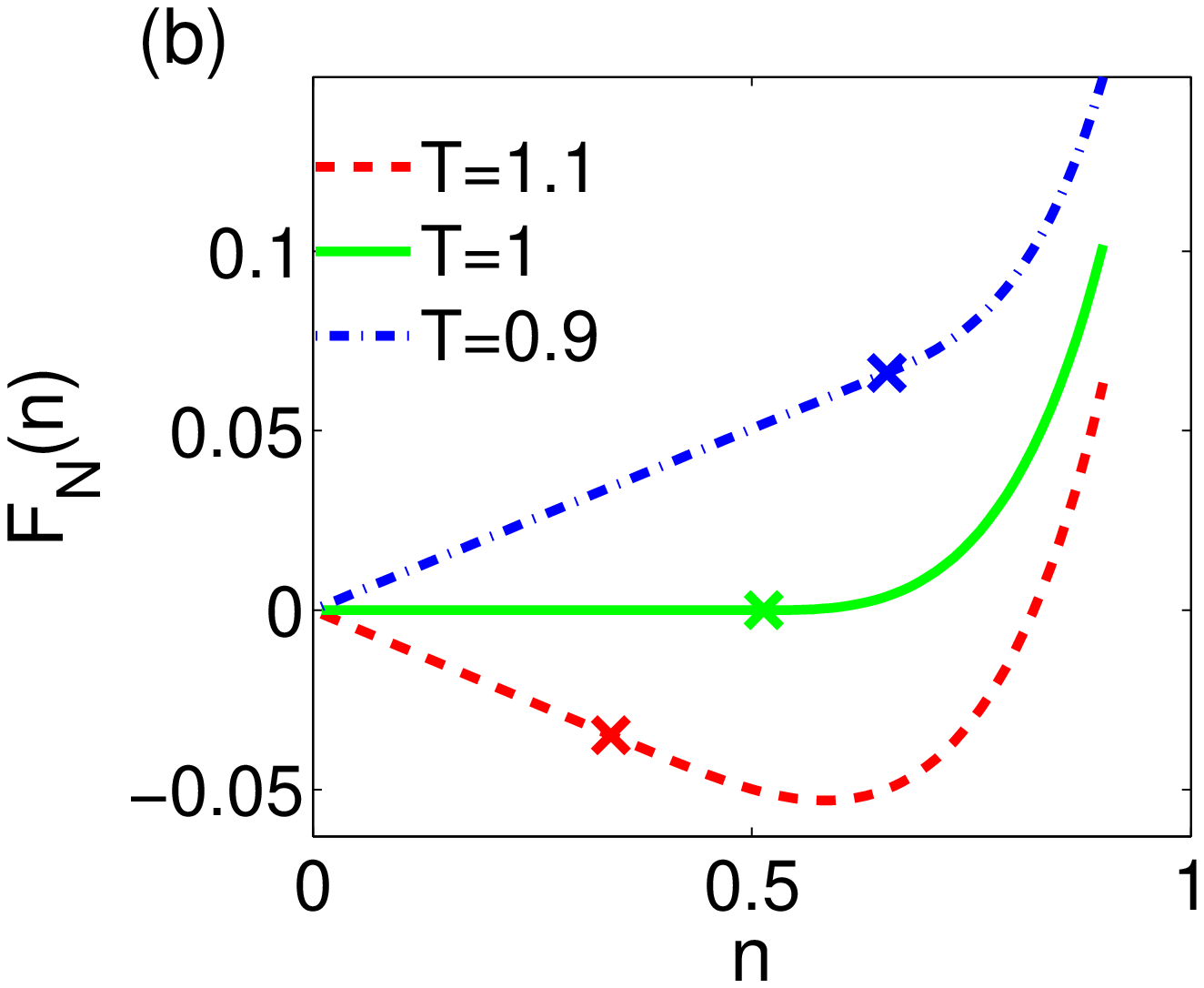}

\caption{\label{fig:LDF} (color online) Free energies of the (a) constant
magnetization ensemble, and (b) constant number of domains ensemble.
The X marks correspond to $\pm m_{c}$ in (a) and $n_{c}$ in (b),
as defined in the text. Notice that in (a) for $T=1.4$ there is no
X mark as $m_{c}=1$, and for $T=1$ there is a single X mark as $m_{c}=0$.
The parameters used to produce these figures and the following ones
are $c=2.5$ and $\Delta=\log\zeta_{c}\approx0.29$ so that $T_{c}=1$.}
\end{figure}

The LDF for the magnetization in zero external magnetic field is given by

\begin{eqnarray}
F_{M}\left(m;\beta\right) & \equiv & -\lim_{L\rightarrow\infty}\frac{1}{L}\log Z_{M}\left(L,mL;\beta\right)\nonumber \\
 & = & \left(\frac{1+m}{2}\right)\log\left[z_{+}^{*}\right]+\left(\frac{1-m}{2}\right)\log\left[z_{-}^{*}\right],\label{eq_res:FM}
\end{eqnarray}
where, for $m\ge0$, $z_{\pm}^{*}$ are given by
\begin{eqnarray}
z_{-}^{*} & = & W_{-}\left(z_{+}^{*},\beta\right),\label{eq_res:zm}\\
z_{+}^{*} & = & \cases{
W_{+}\left(m,\beta\right) & $m<m_{c}$\\
1 & $m\ge m_{c}$ \\ },\label{eq_res:zp}
\end{eqnarray}
with $W_-$ and $W_+$ are given implicitly by Eq. (\ref{eq_C:FM_pole}) and Eq.(\ref{eq_C:fm_sp}) respectively. Since $F_{M}$ is symmetric in $m$, this sets its value
also for $m<0$. Here $W_{\pm}$ are analytic (and non-constant) functions
of their arguments, and $W_{+}\left(m_{c},\beta\right)=1$, where
$m_{c}=1$ for $\beta C\le2$, while it is a decreasing function of
$\beta$ which vanishes at $\beta_{c}$ given by Eq.(\ref{eq:results_betac}).
This implies that if $c\equiv\beta_{c}C>2$, then for a given $\beta$
in the interval $\left[\frac{2}{C},\beta_{c}\right]$, the LDF is
linear for $\left|m\right|>m_{c}$, as can be seen in Fig.\ref{fig:LDF}a.
This linearity is the usual Maxwell construction of a first order
transition in which under a magnetization constraint, the system phase
separates and hence the free energy is essentially the weighted sum
of the corresponding free energies of the phases. In this case the
phases are the normal gas of domains with $\left|m\right|=m_{c}$
and the condensate with $\left|m\right|=1$. There are obviously two
symmetric condensates, with $m=\pm1$, and well below the critical
temperature, when the gas phase becomes unstable, the coexistence
is between those two condensates, and hence the free energy is completely
flat. Note that detailed characterization of the nature of the phase transition requires the knowledge of the $L^{-1}$ correction to the free energy, which will be presented below.
%%Note that while the discontinuity
%in $\left\langle m\right\rangle $ is evident from Fig.\ref{fig:LDF}a,
%in order to understand the origin of the other properties of the transition,
%namely the divergence of the length scale and susceptibility, the
%$L^{-1}$ corrections are crucial.

The LDF of the domains density is given by
\begin{eqnarray}
F_{N}\left(n;\beta\right) & \equiv & -\lim_{L\rightarrow\infty}\frac{1}{L}\log Z_{N}\left(L,nL;\beta\right)\nonumber \\
 & = & \log z^{*}-n\log\left[e^{-\beta\Delta}\Phi_{\beta C}\left(z^{*}\right)\right],\label{eq_res:FN}
\end{eqnarray}
where $z^{*}$ is given by
\begin{equation}
z^{*}=\cases{
W_{n}\left(n,\beta\right) & $n>n_{c}$\\
1 & $n\le n_{c}$ \\ },\label{eq_res:zstar}
\end{equation}
with $W_n$ is given implicitly by Eqs.(\ref{eq_C:FN_saddlepoint}). Here $W_{n}$ is an analytic function of its arguments
such that $W_{n}\left(n,\beta\right)\le1$ for $n>n_{c}$, and it
is an increasing function of $n$ so that equality is achieved only
for $n=n_{c}$. In addition, $n_{c}=0$ for $\beta C\le2$, it is an
increasing function of $\beta$ and it tends to $1$ for $\beta\rightarrow\infty$.
Thus, as for $F_{M}$, we find that $F_{N}$ has a linear part for
$0<n<n_{c}$ and $c>2$. It can also be seen that at the critical temperature,
given by (\ref{eq:results_betac}), the slope of the linear part of
$F_{N}$ vanishes, while it is negative for $\beta<\beta_{c}$ and
positive for $\beta>\beta_{c}$ (see Fig.\ref{fig:LDF}b). This implies
that the minimum of the free energy is at $n^{*}>n_{c}$ for $\beta<\beta_{c}$,
at $n^{*}=0$ for $\beta>\beta_{c}$, and is degenerate on the interval
$\left[0,n_{c}\right]$ for $\beta=\beta_{c}$. Hence for $c>2$ we
find indeed a discontinuous change of $\left\langle n\right\rangle $
at the transition, while for $c\le2$ the change is continuous as
$n_{c}=0$ at the transition.

\paragraph{Finite size corrections}

\begin{figure}
\includegraphics[scale=0.5]{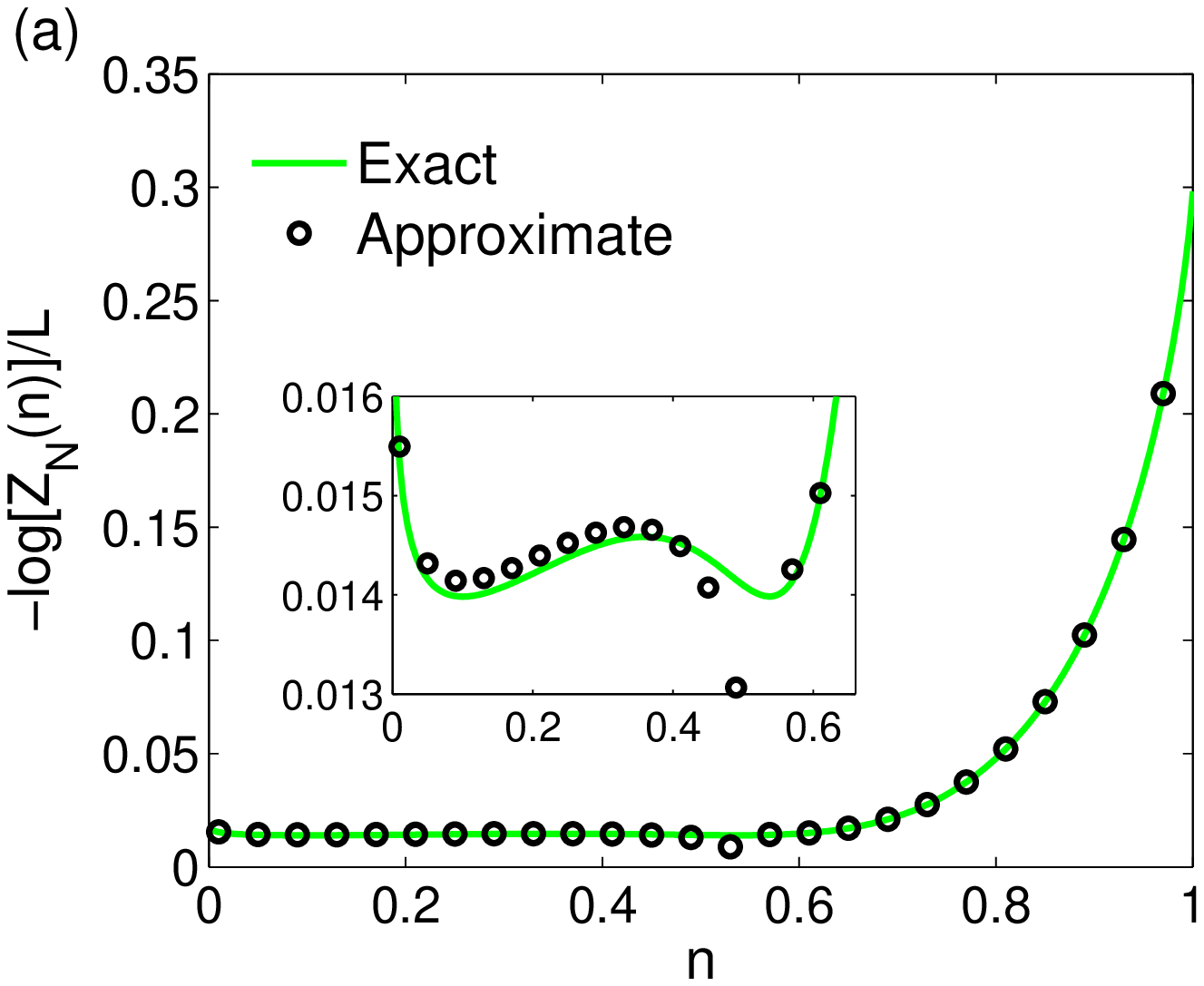}\includegraphics[scale=0.5]{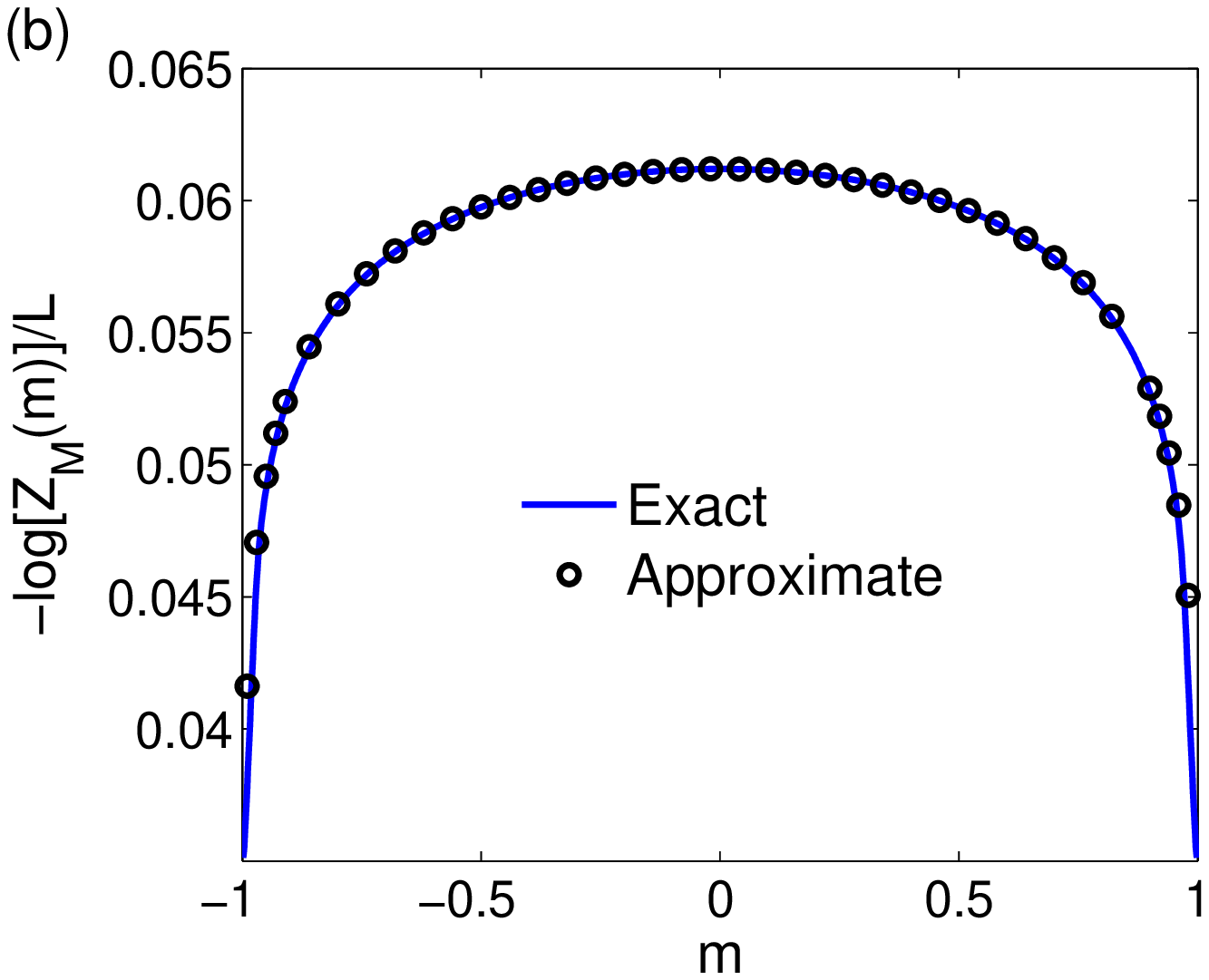}

\caption{\label{fig:LDF_finite}(color online) Exact numerical evaluation ($L=1000$)
vs. analytical approximation (a) for $-\frac{1}{L}\log Z_{N}$ slightly
below $T_{c}$ ($T=0.95T_{c}$), and (b) for $-\frac{1}{L}\log Z_{M}$
for $T=0.5T_{c}$.}
\end{figure}

Fig.\ref{fig:LDF_finite} presents an exact numerical calculation
(see Appendix D) and analytic approximation (see section \ref{sub:Finite-size-corrections})
to $-\frac{1}{L}\log Z_{N}$ and $-\frac{1}{L}\log Z_{M}$ for a finite
system of size $L=1000$. Fig.\ref{fig:LDF_finite}a shows that the
transition is indeed first-order-like, with the usual picture of two
competing wells, and that the linearity of the LDF comes from a Maxwell
construction. In both figures \ref{fig:LDF_finite}a-b it can be seen
that the approximations Eq.(\ref{eq:finit_size_ZN1}-\ref{eq:finit_size_ZM})
are quite accurate. An immediate application of having an explicit approximation
is the observation that the free energy barriers, which suppress fluctuations
of the order parameters, are logarithmic in $L$ and $n$. Hence fluctuations
are distributed following a power law, which implies divergence of
the correlation length and criticality.

\begin{figure}
\begin{centering}
\includegraphics[scale=0.4]{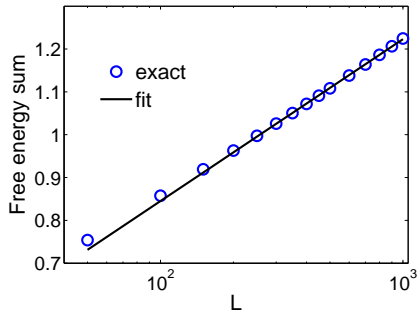}
\par\end{centering}

\caption{\label{fig:finite_size_SN}Scaling of $R(L)$ with $L$. The linearity
of the graph on a semi-logarithmic scale shows the logarithmic scaling,
i.e. $R(L)\sim\log L$.}
\end{figure}

This analytical observation is verified by an exact numerical analysis presented
in Fig.\ref{fig:finite_size_SN}, in which the weight of fluctuations
of the density of domains $n$ at $T_{c}$ are quantified. This is
done by calculating the sum of the free energy of such fluctuations
from $n=0$ to the minimum of the free energy at $n^{*}$
\begin{equation}
R(L)\equiv\int_{0}^{n^{*}}-\log Z_{N}\left(L,Ln;\beta\right)dn.\label{eq_res:R}
\end{equation}
From Eq.(\ref{eq:finit_size_ZN1}-\ref{eq:finite_size_ZN2}) it is evident that there
is a unique minimum of the free energy at $n^{*}(L)>n_{c}$, which
represents a gas of domains. In Fig.\ref{fig:finite_size_SN} it is
shown that $R(L)$ scales logarithmically with $L$,
implying logarithmic free energy barriers.

\subsection{Comparison with the PS and IDSI models}

As was shown above, the TIDSI model can be considered as a symmetric
version of the PS model or as a truncated version of the IDSI model.
In this section we discuss the similarities and differences in the
phenomenology of the three models.

\subsubsection{PS model}

The TIDSI model has a very similar phase diagram to that of the PS model,
as can be appreciated by comparing Fig.\ref{fig:PS_PT} and Fig.\ref{fig:Phase_diagram}a,
recalling the inverted role of the temperature in the two models.
Essentially the mechanism of the phase transition is the same, i.e.
a condensation transition in which a single domain - or a single loop
- becomes macroscopic and encompasses the whole system. There are
a few differences, though, which we shall discuss now.

The most important difference is that at $h=0$, the TIDSI model exhibits
an extreme transition in magnetization, from $m=0$ to $m=\pm1$.
This also happens for $1<c\le2$, for which the PS model exhibits
a second order transition. The reason for this difference is the symmetry
between $+$ and $-$ spins in the TIDSI model, which is lacking in
the PS model. Due to this symmetry, the magnetization vanishes above
the transition (as long as there is no magnetic field). To break the
symmetry there must be either an external symmetry breaking field
or a spontaneous symmetry breaking transition. Note that once a symmetry
breaking field is applied, the transition becomes almost identical
to the PS transition, as for $1<c(h)\le2$ the transition is second
order. A second related difference is the existence of two order parameters
which behave differently for $1<c\le2$: the density of domains, $n$,
which is continuous and the magnetization which is discontinuous.
This has to do with the symmetry of the model under magnetization reversal,
which has no counterpart for the $n$ order parameter. Hence the behavior of $n$
in the TIDSI model is the same as that in the PS model.

Another difference has to do with the parameter $c$. In the PS model
this parameter is the universal exponent controlling the number of
configurations of a self-avoiding random loop. The value of $c$ is
hence independent of geometry and depends only on the spatial dimension and
topological constraints. In the TIDSI model, on the other hand, the
parameter $C$ itself is rather arbitrary, and moreover the value
of the critical temperature $\beta_{c}$ depends both on $C$ and
on $\Delta$, or the nearest neighbors interactions. Hence $c\equiv\beta_{c}C$
depends on the model parameters rather than being universal.

\subsubsection{IDSI model}

The main similarity between the IDSI model and the TIDSI model is
that both exhibit a mixed order symmetry breaking phase transition.
There are, however, some distinctions between the two models. While the TIDSI
has a finite magnetic field transition, the IDSI model exhibits no
such transition, as expected from Ising ferromagnet with two body
interactions by the Lee-Yang theorem \cite{lee1952statistical,blythe2003lee,fisher1982scaling}.
Another difference is the divergence of length-scale at the transition,
which is algebraic for the TIDSI model and essential singularity in
the case of the IDSI model. The origin of this behavior is discussed
in section \ref{sec:RG-analysis} through the renormalization group
analysis.

Perhaps a more fundamental difference between the models is the fact
that in the IDSI there is no formation of a macroscopic domain (where
a domain is defined as a consecutive set of spins all having the same
sign). While the density of domains was not analyzed in the case of
IDSI, it seems that it does not vanish for any finite temperature,
though an interesting question is whether it is analytic at the transition
or not. The lack of condensation in the IDSI is obviously the reason
that the Thouless effect in this model is not extreme, i.e. that the
magnetization jumps to a finite number $<1$.

\section{\label{sec:Grand-canonical-analysis}Grand canonical analysis}

We now establish analytically the results stated above. The model
(\ref{eq_model:ttt_Hamiltonian2}) is defined in the canonical ensemble
where the chain length $L$ is fixed. In an ensemble where $L$, $N$
and $M$ are fixed the partition function and the associated free
energy (or large deviations function) are given by
\begin{eqnarray}
\fl Z_{0}\left(L,M,N;\beta\right)  = e^{-\beta\Delta N}\sum_{l_{1}=1}^{\infty}...\sum_{l_{N}=1}^{\infty}\prod_{a=1}^{N}l_{a}^{-\beta C} I\left(L=\sum l_{a}\right) \times \nonumber \\
\qquad \qquad I\left(M=-\sigma_{1}\sum_{a=1}^{N}\left(-1\right)^{a}l_{a}\right),\label{eq_GC:Z_0}\\
\fl F_{0}\left(M,N;\beta\right)  =  \lim_{L\rightarrow\infty}\frac{1}{L}\log Z_{0}\left(L,M,N;\beta\right)
\end{eqnarray}
As common in statistical mechanics, instead of evaluating the constrained
partition sum (\ref{eq_GC:Z_0}) itself, we calculate the generating
function
\begin{equation}
Q\left(p,h,\mu;\beta\right)=\sum_{L,M,N}Z_{0}\left(L,M,N;\beta\right)e^{L\beta p}e^{\beta hM}e^{\beta\mu N}.\label{eq_GC:Q_def}
\end{equation}
For simplicity we assume symmetric boundary conditions with
\begin{equation}
\sigma_{1}=1\qquad;\qquad\sigma_{L}=-1,\label{eq_GC:BC}
\end{equation}
which imply an even number of domains. Eqs.(\ref{eq_GC:Z_0}-\ref{eq_GC:BC})
yield
\begin{eqnarray}
Q\left(p,h,\mu;\beta\right) & = & \sum_{N}e^{-\beta\left(\Delta-\mu\right)N}\prod_{a=1}^{N}\left[\sum_{l=1}^{\infty}\frac{e^{\beta pl}}{l^{\beta C}}\exp\left(\left(-1\right)^{a}\beta hl\right)\right]\nonumber \\
 & = & \frac{e^{2\beta\left(\mu-\Delta\right)}U\left(p+h\right)U\left(p-h\right)}{1-e^{2\beta\left(\mu-\Delta\right)}U\left(p+h\right)U\left(p-h\right)},\label{eq_GC:Q_res}
\end{eqnarray}
with
\begin{equation}
U\left(x\right)=\sum_{l}\frac{e^{\beta xl}}{l^{\beta C}}=\Phi_{\beta C}\left(e^{\beta x}\right).\label{eq_GC:U_def}
\end{equation}
The function $\Phi_{\gamma}\left(u\right)$ is the polylogarithm function,
which has the following properties:
\begin{enumerate}
\item $\Phi_{\gamma}\left(u\right)$ is analytic in the complex plane except
for a branch-cut along $[1,\infty)$
\item $\Phi_{\gamma}\left(1\right)=\infty$ for $\gamma\le1$ and $\Phi_{\gamma}\left(1\right)=\zeta(\gamma)$
for $\gamma>1$ where $\zeta\left(\gamma\right)$ is the Riemann Zeta
function.
\item $\frac{d}{dx}\Phi_{\gamma}\left(e^{x}\right)=\Phi_{\gamma-1}\left(e^{x}\right)$
\item Expanding around $x=0$, $\Phi_{\alpha}\left(e^{x}\right)=\Gamma(1-\alpha)\left(-x\right)^{\alpha-1}+\sum_{k=0}^{\infty}\frac{\zeta\left(s-k\right)}{k!}x^{k}$
, where $\Gamma\left(\right)$ is the Gamma function.
\end{enumerate}
We proceed by using (\ref{eq_GC:Q_res}) to derive the results discussed
in section \ref{sec:The-model}, namely the discontinuity of the magnetization,
the divergence of the correlation length etc. Then we argue that the
low temperature phase is characterized by a single macroscopic domain
- a \emph{condensate}. Finally we justify the derivation for the low
temperature phase using a regularization argument.

\subsection{The normal phase}

The thermodynamic behavior of the system is determined by the grand
potential \cite{fisher1984walks,bar2014mixed}
\begin{equation}
p^{*}=\min_{M,N}\left\{ F_{0}\left(M,N\right)-h\frac{M}{L}-\mu\frac{N}{L}\right\} ,
\end{equation}
The average of the order parameters $m\equiv M/L$ and $n\equiv N/L$
are given by
\begin{equation}
\left\langle m\right\rangle =-\frac{\partial p^{*}}{\partial h}\qquad;\qquad\left\langle n\right\rangle =-\frac{\partial p^{*}}{\partial\mu}.
\end{equation}
According to Eq.(\ref{eq_GC:Q_def}), $e^{\beta p^{*}}$ corresponds
to the singularity of $Q$ closest to the origin (most negative $p^{*}$),
as it is the radius of convergence of its defining series. Inspecting
(\ref{eq_GC:Q_res}) we see that the singularity can stem either from
the denominator becoming $0$, i.e.
\begin{equation}
U\left(p^{*}+h\right)U\left(p^{*}-h\right)=e^{2\beta\left(\Delta-\mu\right)},\label{eq_GC:sing_1}
\end{equation}
or from the branch point of $U$, i.e.
\begin{equation}
p^{*}+\left|h\right|=0.\label{eq_GC:sing_2}
\end{equation}
In the analysis below we focus on the regime of small $\mu$. For
high temperatures, $\beta\rightarrow0$, $p^{*}$ is the solution of
(\ref{eq_GC:sing_1}). By differentiating (\ref{eq_GC:sing_1}) with
respect to $h$ we find
\begin{eqnarray}
m(h)  =  \frac{\Psi_{+}\left(p^{*},h\right)-\Psi_{-}\left(p^{*},h\right)}{\Psi_{+}\left(p^{*},h\right)+\Psi_{-}\left(p^{*},h\right)},\label{eq_GC:M_h_highT}\\
\Psi_{\pm}(p,h)  \equiv  U'\left(p\pm h\right)U\left(p\mp h\right),
\end{eqnarray}
and hence $m\rightarrow0$ for $h\rightarrow0$ (as $\Psi_{+}\left(p,0\right)=\Psi_{-}\left(p,0\right)$).
However, as $\beta$ increases, the RHS of (\ref{eq_GC:sing_1}) increases
(for $\Delta>\mu$) while the LHS decreases (for a given $p^{*}$)
and hence $p^{*}$ is an increasing function of $\beta$. Therefore
there is a critical value of $\beta$ such that Eq.(\ref{eq_GC:sing_2})
is also satisfied. This happens at $\beta_{c}$ which is the solution
of
\begin{equation}
\Phi_{\beta_{c}C}\left(1\right)\Phi_{\beta_{c}C}\left(e^{-2\beta\left|h\right|}\right)=e^{2\beta_{c}\left(\Delta-\mu\right)}.\label{eq_GC:beta_c}
\end{equation}
From (\ref{eq_GC:beta_c}) it is clear that the parameter $c\equiv\beta_{c}C$
satisfies $c>1$ since $\Phi_{c}(1)$ diverges for $c\le1$. As is
described in section \ref{sub:Condensate-phase--}, below the transition,
namely for $\beta>\beta_{c}$, the singularity of $Q$ closest to
the origin is given by Eq.(\ref{eq_GC:sing_2}), i.e. $p^{*}=-\left|h\right|$,
and hence
\begin{equation}
m(h)=sign(h)\equiv\cases{
1 & $h>0$\\
-1 & $h<0$ \\}.\label{eq_GC:M_h_lowT}
\end{equation}
Eq.(\ref{eq_GC:M_h_lowT}) then proves the existence of a discontinuity
of the magnetization, where $m(h\rightarrow0)$ jumps from $0$ to
$\pm1$ at some finite $\beta_{c}$. In addition, we see from (\ref{eq_GC:M_h_highT}-\ref{eq_GC:M_h_lowT})
that there is also a phase transition for $h\neq0$, since while for
$T>T_{c}$ the magnetization is given by (\ref{eq_GC:M_h_highT})
so that $0<\left|m\left(h\right)\right|<1$, for $T<T_{c}$ the magnetization
is $m\left(h\right)=sign(h)$. The order of this transition depends
on $U'\left(0\right)$: When it diverges the transition is continuous
as can be verified from Eq.(\ref{eq_GC:M_h_highT}), while when it
is finite the transition involves a discontinuity of the magnetization.
From the properties of the polylogarithm function this implies that
the magnetization is continuous for $c\left(h\right)\equiv\beta_{c}\left(h\right)C\le2$
and discontinuous for $c\left(h\right)>2$.

Calculating the average $N$ in the high temperature regime by differentiating
(\ref{eq_GC:sing_1}) with respect to $\mu$ yields
\[
n\left(h\right)=\frac{2\beta e^{2\beta\left(\Delta-\mu\right)}}{\Psi_{+}\left(p^{*},h\right)+\Psi_{-}\left(p^{*},h\right)}.
\]
 In the low temperature phase, where $p^{*}=-\left|h\right|$, the
result is just $n=0$. Hence $n$ is continuous through the transition
if $U'(0)$ diverges (and therefore also $\Psi_{+}\left(-\left|h\right|,h\right)$
or \textbf{$\Psi_{-}\left(-\left|h\right|,h\right)$ }diverge\textbf{)
}, i.e. $c\le2$, and is discontinuous otherwise. Thus the two order
parameters $m$ and $n$ behave differently at $h=0$. While $m$
is discontinuous at the transition for any $c$, $n$ is continuous
for $1<c\le2$ and discontinuous only for $c>2$. On the other hand
on the $h\neq0$ transition lines both $m$ and $n$ change continuously
for $c\left(h\right)\le2$ and jump for $c\left(h\right)>2$.

One can also use the above results to calculate the magnetic susceptibility
$\chi\equiv\frac{\partial m}{\partial h}$ and the distribution of
domain sizes $P(l)$, which defines a typical length scale which diverges
at the transition. Differentiating (\ref{eq_GC:M_h_highT}) with respect
to $h$ yields
\begin{equation}
\fl\chi=\frac{2\left(\Psi_{-}\partial_{h}\Psi_{+}-\Psi_{+}\partial_{h}\Psi_{-}\right)\left(\Psi_{+}+\Psi_{-}\right)+2\left(\Psi_{+}-\Psi_{-}\right)\left(\Psi_{+}\partial_{p}\Psi_{-}-\Psi_{-}\partial_{p}\Psi_{+}\right)}{\left(\Psi_{+}+\Psi_{-}\right)^{3}}.
\end{equation}
As $\Psi_{\pm}$ involve $U'\left(p^{*}\pm h\right)$, $\chi$ involves
$U''\left(p^{*}\pm h\right)$. It is easy to see that there is no
cancellation of these terms, and hence if $U''(0)$ diverges then
$\chi$ diverges. From the properties of the Polylog it is evident that $U''(0)$ diverges if $c\le3$.

Finally, the distribution of the size of $+$ and $-$ domains is
given by
\begin{eqnarray}
P_{\pm}\left(l\right) & \simeq & \frac{Z_{C}\left(L-l,h\right)}{Z_{C}\left(L,h\right)}\times\frac{e^{\pm\beta hl}}{l^{\beta C}}\nonumber \\
 & = & \frac{e^{-\beta\left(p^{*}\pm h\right)l}}{l^{\beta C}}=\frac{e^{-l/\xi_{\pm}}}{l^{\beta C}},\label{eq_GC:domain_size_dist}
\end{eqnarray}
where we used $Z_{C}\sim e^{-L\beta p^{*}}$ and defined $\xi_{\pm}\equiv\left[\log\left(p^{*}\pm h\right)\right]^{-1}$.
The length scales $\xi_{\pm}$ are not exactly the correlation length
of the spin-spin correlation function, but they are lower bounds for
it. For $h>0$ ($h<0$) the length scale $\xi_{+}$ ($\xi_{-}$) diverges
as $T\rightarrow T_{c}$ for all $c$, which implies that the correlation
length diverges as well. Expanding Eq.(\ref{eq_GC:sing_1}) near the
transition, i.e. $t\equiv T-T_{c}\ll1$, $\delta p\equiv-\left|h\right|-p^{*}\ll1$,
and using property 4 of the polylogarithm function we get
\[
t\sim\left(\delta p\right)^{\gamma}\quad;\quad\gamma=\min\left(c-1,1\right).
\]
The algebraic divergence of the correlation length Eq.(\ref{eq_res:xi}) directly follows
from this relation.

\subsection{Appearance of condensate in the low temperature phase}

In the low temperature phase, where $n=0$, the number of domains
in the system is sub-extensive. We argue that in fact this phase is
composed of a single macroscopic domain by showing that this state
is more favorable than having two condensates. The argument is similar
to that given in \cite{evans2005nonequilibrium} for condensation
in the zero range process.

For a system with a single macroscopic domain and the boundary conditions
(\ref{eq_GC:BC}), the partition sum scales as $Z_{1}\sim L^{-\beta C}$.
For a system with two macroscopic domains the partition function $Z_{2}$
is a sum of $O\left(L\right)$ terms, each represents a different
location of the domain wall. Each term scales as $\left(L^{-\beta C}\right)^{2}$
and hence $Z_{2}\sim L^{1-2\beta C}$. In the low temperature phase
$\beta C>1$ and hence $-\beta C>1-2\beta C$ and the single condensate
state is preferable. This argument can be easily extended to
other configurations of condensates. From the fact that there is a
single condensate, one can easily deduce that $m=\pm1$ in the low
temperature phase.

\subsection{Condensate phase - regularization\label{sub:Condensate-phase--}}

The above argument for the relation $p^{*}=-\left|h\right|$ in the low temperature phase is not mathematically justified, as the sum defining the grand partition function (\ref{eq_GC:Q_def}) is not well behaved in this regime.

%%The validity of the relation $p^{*}=-\left|h\right|$ (Eq.(\ref{eq_GC:sing_2}))
%%in the low temperature phase is not clear from the above derivation,
%%as the grand partition function (\ref{eq_GC:Q_def}) is not analytic
%%at this point.
One way to justify it is to invert the z-transform
(\ref{eq_GC:Q_def}), thus finding $Z_{0}$ directly and validating
(\ref{eq_GC:sing_2}) for $T<T_{c}$. This route is taken in section
\ref{sec:Canonical-analysis}. Here we follow a different procedure,
whereby the grand-canonical ensemble is regularized by introducing an
upper cutoff to the domain length, thus making all quantities analytic,
and then taking the upper cutoff to infinity. This procedure is well
suited for models with a condensation phenomena (similar to BEC) and
it was used in such context \cite{kafri2002melting,evans2005nonequilibrium,bar2011denaturation}.

Let us consider the TIDSI model as defined in Eq.(\ref{eq_model:ttt_Hamiltonian2})
and introduce an upper cutoff $\Lambda$ on the domain length. The
partition function for this modified model is

\begin{eqnarray}
Q_{\Lambda}\left(p,h,\mu;\beta\right) & = & e^{\beta\Delta}\sum_{N}e^{-\beta\left(\Delta-\mu\right)N}\prod_{a=1}^{N}\left[\sum_{l=1}^{\Lambda}\frac{e^{\beta pl}}{l^{\beta C}}\exp\left(\left(-1\right)^{a}\beta hl\right)\right]\nonumber \\
 & = & e^{\beta\Delta}\frac{e^{2\beta\left(\mu-\Delta\right)}U_{\Lambda}\left(p+h\right)U_{\Lambda}\left(p-h\right)}{1-e^{2\beta\left(\mu-\Delta\right)}U_{\Lambda}\left(p+h\right)U_{\Lambda}\left(p-h\right)},\label{eq_GC:Q_res-1}
\end{eqnarray}
with
\begin{equation}
U_{\Lambda}\left(x\right)  =  \sum_{l=1}^{\Lambda}\frac{e^{\beta xl}}{l^{\beta C}}=\Phi_{\beta C}^{\Lambda}\left(e^{\beta x}\right).
\end{equation}
The function $\Phi_{\gamma}^{\Lambda}\left(u\right)$ is a truncated
version of the polylogarithm, which is analytic for any $u$ and $\gamma$.
For given $h$ and $\mu$ the thermodynamic limit $L\rightarrow\infty$
is again given by the most negative singularity of $Q_{\Lambda}$,
i.e. by the solution of
\[
U_{\Lambda}\left(p^{*}+h\right)U_{\Lambda}\left(p^{*}-h\right)=e^{2\beta\left(\Delta-\mu\right)}.
\]
Since $U_{\Lambda}$ is an analytic function, this equation has a
solution for all $\beta$. Like before, $p^{*}$ is an increasing
function of $\beta$ and therefore there is a temperature $\beta_{c}\left(\Lambda\right)$
for which $p^{*}\left(h,\mu,\beta_{c}\left(\Lambda\right)\right)=-\left|h\right|$.
For $\beta>\beta_{c}\left(\Lambda\right)$ one has $p^{*}>-\left|h\right|$.
In the limit $\Lambda\rightarrow\infty$, $\beta_{c}\left(\Lambda\right)\rightarrow\beta_{c}$
as given by Eq.(\ref{eq_GC:beta_c}). For any $u>1$, $\lim_{\Lambda\rightarrow\infty}\Phi_{\beta C}^{\Lambda}(u)=\infty$,
and hence for $\beta>\beta_{c}$ and in the limit $\Lambda\rightarrow\infty$
the solution satisfies $p^{*}\rightarrow-\left|h\right|$ thus validating
(\ref{eq_GC:sing_2}).

\section{\label{sec:Canonical-analysis}Canonical analysis}

The above analysis proves the existence of the transition and its
unique properties. To get deeper understanding of the mechanism of
the condensation transition it is instructive to study the model within
the canonical ensemble. Critical phenomena are commonly described
by Landau theory of phase transitions, which provides more details
on the nature of such transitions. While the Landau theory is usually
based on phenomenological analysis, in our case the model can be solved
exactly and hence we can calculate the large deviation functions
(LDF) $F_{M}$ and $F_{N}$ and their finite size corrections, which
play the role of Landau free energy in this analysis. It turns out that unlike the basic assumption of the Landau theory, namely that the free energy is an analytic function of the order parameter, the functions $F_{M}$ and $F_{N}$ turns out to be non-analytic in $n$ and $m$ respectively.

\subsection{Large deviations functions}

To find the LDF and their finite size correction at zero external field we invert the $z$-transform
(\ref{eq_GC:Q_def}) using the Cauchy integral formula, and use complex
analysis techniques to evaluate the result to relevant order.

\subsubsection{Magnetization large deviations fuction $F_{M}$}

\begin{figure}
\includegraphics[scale=0.4]{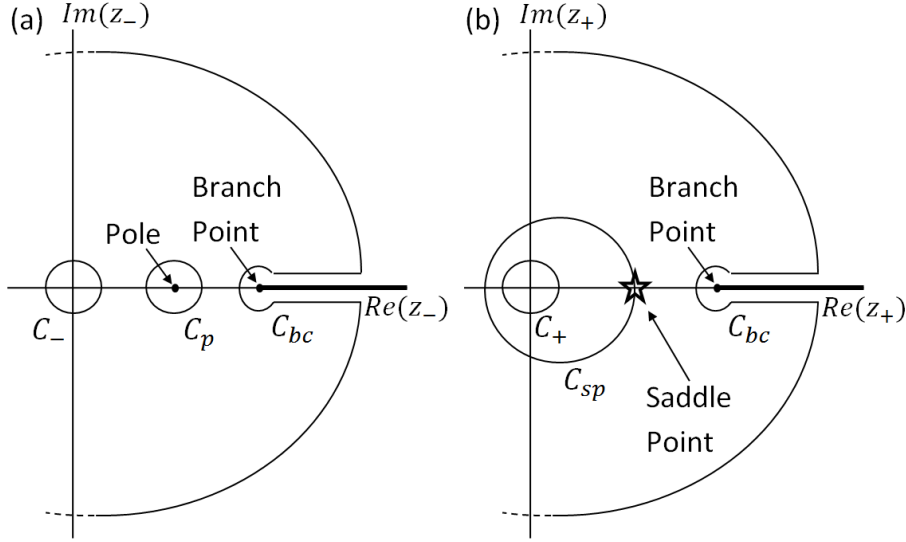}

\caption{\label{fig:Countors}Contours of integration used in the calculations
of $Z_{M}$ and $Z_{N}$ (see text).}
\end{figure}

For ease of notations we define
\begin{equation}
L_{\pm}\equiv\frac{1}{2}\left(L\pm M\right)\quad;\quad z_{\pm}\equiv e^{\beta\left(p\pm h\right)}\quad;\quad A\equiv e^{-\beta\Delta}.\label{eq:FM_notations}
\end{equation}
We carry out the analysis for $M\ge0$ (or $L_{+}\ge L_{-}$). The LDF for $M<0$
is obtained from that of $M>0$ by symmetry. The partition function
$Z_{M}$ for the ensemble in which $L$ and $M$ (but not $N$) are
fixed is given then by
\begin{eqnarray}
Z_{M}\left(L,M;\beta\right) & = & \oint_{C_{+}}\frac{dz_{+}}{2\pi i}\oint_{C_{-}}\frac{dz_{-}}{2\pi i}\frac{\tilde{Q}\left(z_{+},z_{-};\beta\right)}{z_{+}^{L_+ + 1}z_{-}^{L_- +1}}\label{eq_C:Z_M},\\
\tilde{Q}\left(z_{+},z_{-};\beta\right) & \equiv & Q\left(\frac{1}{\beta}\log\left(z_{+}z_{-}\right),\frac{1}{\beta}\log\left(z_{+}/z_{-}\right),0;\beta\right).\label{eq_C:Q_tilde}
\end{eqnarray}
Using the explicit form of $Q$ (\ref{eq_GC:Q_res}) we get

\[
Z_{M}\left(L,M;\beta\right)=\frac{A}{\left(2\pi i\right)^{2}}\oint\frac{dz_{+}dz_{-}}{z_{+}^{L_{+}+1}z_{-}^{L_{-}+1}}\frac{\Phi_{\beta C}\left(z_{+}\right)\Phi_{\beta C}\left(z_{-}\right)}{1-A^{2}\Phi_{\beta C}\left(z_{+}\right)\Phi_{\beta C}\left(z_{-}\right)}.
\]
Carrying out the integration over $z_{-}$ we note that for high enough
temperatures the singularity closest to the origin in the $z_{-}$
plane is a simple pole at $z_{-}^{*}\left(z_{+}\right)$ which satisfies
\begin{equation}
\Phi_{\beta C}\left(z_{+}\right)\Phi_{\beta C}\left(z_{-}^{*}\right)=A^{-2}.\label{eq_C:FM_pole}
\end{equation}
This is the same equation as (\ref{eq_GC:sing_1}). Hence the contour
of integration $C_{-}$ can be deformed to a contour encircling this
pole $C_{p}$, plus a contour with a larger radius $C_{bc}$(see Fig.\ref{fig:Countors}a).
Due to the $z_{-}^{-L_{-}}$ factor only the pole contour contributes,
and hence
\begin{eqnarray}
\fl Z_{M}\left(L,M;\beta\right) & \approx & \frac{1}{2\pi iA^{3}}\oint_{C_{+}}\frac{dz_{+}}{z_{+}^{L_{+}+1}}\frac{\left(z_{-}^{*}\right)^{-L_{-}}}{\Phi_{\beta C-1}\left(z_{+}\right)\Phi_{\beta C}\left(z_{-}^{*}\right)}\nonumber \\
\fl & \equiv & \frac{1}{2\pi iA^{3}}\oint dz_{+}e^{-Lf_{m}\left(m,z_{+};\beta\right)},\label{eq_C:FM_fmz_def}\\
\fl f_{m}\left(m,z_{+};\beta\right) & = & \left(\frac{1+m}{2}\right)\log\left[z_{+}\right]+\left(\frac{1-m}{2}\right)\log\left[z_{-}^{*}\right]+O\left(\frac{1}{L}\right).\label{eq_C:FM_fmz_res}
\end{eqnarray}
We now proceed to carry out the integration over $z_{+}$. In the
$z_{+}$ plane there is a branch-cut for $z_{+}\in[1,\infty)$ due
to the polylogarithm function. If $f_{m}\left(m,z_{+};\beta\right)$ has an
extremum for $\left|z_{+}\right|<1$, the saddle point method can
be applied by deforming the contour $C_{+}$ to a contour $C_{sp}$
which passes through the saddle point as in Fig.\ref{fig:Countors}b,
yielding
\begin{equation}
F_{M}\left(m;\beta\right)=-\lim_{L\rightarrow\infty}\frac{1}{L}\log Z_{M}\left(L,M;\beta\right)=f_{m}\left(m,z_{+}^{*}\left(m;\beta\right);\beta\right),\label{eq_C:FM_highT}
\end{equation}
with the saddle point $z_{+}^{*}$ satisfying
\begin{equation}
0=\frac{d}{dz_{+}}\left.f_{m}\left(m,z_{+};\beta\right)\right|_{z_{+}^{*}}=\frac{1+m}{2z_{+}^{*}}+\left(\frac{1-m}{2z_{-}^{*}}\right)\left.\frac{dz_{-}^{*}}{dz_{+}}\right|_{z_{+}^{*}}.\label{eq_C:fm_sp}
\end{equation}
Eq.(\ref{eq_C:FM_highT}) is the result for the LDF $F_{M}$ in the
high temperature regime. We shall now show that at a certain temperature
there is no longer a saddle point for $\left|z_{+}\right|<1$ and
hence a different approach should be followed. By differentiating Eq.(\ref{eq_C:FM_pole})
with respect to $z_{+}$ and using Eq.(\ref{eq_C:fm_sp}) the saddle point condition reads
\begin{equation}
\frac{\Phi_{\beta C-1}\left(z_{+}^{*}\right)\Phi_{\beta C}\left(z_{-}^{*}\right)}{\Phi_{\beta C}\left(z_{+}^{*}\right)\Phi_{\beta C-1}\left(z_{-}^{*}\right)}=\frac{1+m}{1-m}.\label{eq_C:FM_saddlepoint}
\end{equation}
For fixed $z_{\pm}^{*}$ the LHS of (\ref{eq_C:FM_saddlepoint}) is
a decreasing function of $\beta$, while for fixed $\beta$ it is
an increasing function of $z_{+}$ as proved in Appendix A. The RHS
of (\ref{eq_C:FM_saddlepoint}) is an increasing function of $m$,
and hence $z_{+}^{*}\left(m;\beta\right)$ is an increasing function
of both $\beta$ and $m$. Therefore, for a given $\beta$ such that
$\beta C>2$ there is a critical value of $m$, denoted $m_{c}$,
such that for $m>m_{c}$ there is no saddle point for $\left|z_{+}\right|<1$.
The value of $m_{c}$ is given by
\begin{equation}
\frac{\zeta\left(\beta C-1\right)\Phi_{\beta C}\left(z_{-}^{*}\left(1\right)\right)}{\zeta\left(\beta C\right)\Phi_{\beta C-1}\left(z_{-}^{*}\left(1\right)\right)}=\frac{1+m_{c}}{1-m_{c}}.\label{eq_C:FM_mc}
\end{equation}
If there is no saddle point, one can instead deform the contour $C_{+}$
to contour $C_{bc}$ which wraps the branch cut and close in a large
circle with radius tending to infinity as in Fig.\ref{fig:Countors}.
The details of the calculations are involved and hence they are deferred
to Appendix B. However the result is simple, namely $z_{+}$ is frozen at
$z_{+}=1$ for all $m>m_{c}$, i.e.
\begin{equation}
F_{M}\left(m;\beta\right)=f_{m}\left(m,1;\beta\right)=\left(\frac{1-m}{2}\right)\log\left[z_{-}^{*}\left(1\right)\right].\label{eq_C:FM_lowT}
\end{equation}
Note that the analysis is valid for $m\ge0$. For $m<0$ the roles
of $z_{+}$ and $z_{-}$ should be inverted. Hence we find that for
$m>m_{c}$ the LDF is linear in $m$ as depicted in Fig.\ref{fig:LDF}a.
From Eq.(\ref{eq_C:FM_pole}) it can be seen that $z_{-}^{*}(1)$
is an increasing function of $\beta$, which implies that the slope
of $m$ decreases with $\beta$ (as $z_{-}^{*}<1$). At some critical
value $\beta_{c}$, one has $z_{-}^{*}=1$ and thus $F_{M}$ becomes flat (with
$0$ slope), as this is also the point where $m_{c}=0$. The critical
temperature is given by
\begin{equation}
\zeta\left(\beta_{c}C\right)=e^{\beta_{c}\Delta},\label{eq_C:FM_betac}
\end{equation}
which is the same as Eq.(\ref{eq_GC:beta_c}) for $h=0$. For $\beta>\beta_{c}$
the singularity closest to the origin is no longer the pole (\ref{eq_C:FM_pole})
but the branch-point, and hence $F_{M}=0$ for all $m$.

In summary the free energy $F_{M}\left(m;\beta\right)$ is given by
(\ref{eq_C:FM_highT}) for $\left|m\right|<m_{c}$, and by (\ref{eq_C:FM_lowT})
for $\left|m\right|>m_{c}$, with $m_{c}$ given by (\ref{eq_C:FM_mc})
(see Fig.\ref{fig:LDF}a). The free energy is linear in $m$ for $\left|m\right|>m_{c}$.
Note that Eq.(\ref{eq_C:FM_pole}) defines implicitly the function
$W_{-}\left(z_{+},\beta\right)$ which appears in Eq.(\ref{eq_res:zm}).
Similarly $W_{+}\left(m,\beta\right)$ is defined implicitly by Eq.(\ref{eq_C:FM_saddlepoint})
for $\left|m\right|<m_{c}$ and by $z_{+}=1$ for $\left|m\right|\ge m_{c}$.

The linearity of the LDF for $\left|m\right|>m_{c}$ is a manifestation
of a phase coexistence, i.e. it is the Maxwell construction \cite{Hu1987}.
The coexistence is between a normal phase which consists of microscopic
domains and a phase with a single macroscopic domain. That is, the
most probable way to implement a high magnetization $m>m_{c}$ is
by breaking the system to a ``normal gas'' of domains with total
density $m_{c}$ and a macroscopic condensate with total density $m-m_{c}$.
At the transition the slope of $F_{M}$ vanishes, which implies that
the gas phase and the condensate phase have the same free energy (in
the thermodynamic limit). This description may correspond just as
well to a usual first order phase transition, and hence from this
analysis it is not clear where does the mixed nature of the transition
comes from. As we shall see in section \ref{sub:Finite-size-corrections},
this comes from the logarithmic barriers in $L$ between the phases.

\subsubsection{Domains density large deviations function $F_{N}$}

The analysis for $F_{N}$ is similar to the analysis of $F_{M}$,
so some details will be spared. We start with
\[
Z_{N}\left(L,N;\beta\right)=\frac{A}{\left(2\pi i\right)^{2}}\oint\frac{dzdq}{z^{L+1}q^{N+1}}\frac{q^{2}\Phi_{\beta C}\left(z\right)^{2}}{1-A^{2}q^{2}\Phi_{\beta C}\left(z\right)^{2}},
\]
where $z=e^{\beta p}$ and $q=e^{\beta\mu}$. The pole equation which
corresponds to the integration over $q$ is
\begin{equation}
q^{*}\left(z;\beta\right)=\frac{1}{A\Phi_{\beta C}\left(z\right)}.\label{eq_C:FN_pole}
\end{equation}
Again it is a special case of (\ref{eq_GC:sing_1}) for $h=0$. Carrying
out the pole integral yields
\begin{eqnarray}
Z_{N}(L,N;\beta) & = & \oint\frac{dz}{2\pi i}\frac{\left[A\Phi_{\beta C}\left(z\right)\right]^{Ln}}{Az^{L+1}}\nonumber \\
 & \approx & \oint\frac{dz}{2\pi i}e^{-Lf_{n}\left(n,z;\beta\right)},\label{eq_C:FN_fn_def}\\
f_{n}\left(n,z;\beta\right) & \equiv & \log z-n\log\left[A\Phi_{\beta C}(z)\right].\label{eq_C:FN_fn_res}
\end{eqnarray}
A saddle point of $f_{n}$, when exists, is given by
\begin{equation}
\frac{\Phi_{\beta C-1}\left(z^{*}\right)}{\Phi_{\beta C}\left(z^{*}\right)}=\frac{1}{n}.\label{eq_C:FN_saddlepoint}
\end{equation}
Deforming the contour to pass through it yields
\begin{equation}
F_{N}\left(n;\beta\right)=-\lim_{L\rightarrow\infty}\frac{1}{L}\log Z_{N}\left(L,N;\beta\right)=f_{n}\left(n,z^{*};\beta\right).\label{eq_C:FN_highT}
\end{equation}
The LHS of (\ref{eq_C:FN_saddlepoint}) is a decreasing
function of $\beta$ and an increasing function of $z$, while the
RHS is a decreasing function of $n$, and therefore $z^{*}\left(n,\beta\right)$
is an increasing function of both $\beta$ and $n$. Hence, if $\beta C>2$
there exists $n_{c}$ such that for $n<n_{c}$ there is no saddle
point for $\left|z\right|<1$, and $n_{c}$ is given by
\begin{equation}
n_{c}=\frac{\zeta(\beta C)}{\zeta(\beta C-1)}.\label{eq_C:nc_val}
\end{equation}
Then, by branch-cut integration (Appendix C) we get
\begin{equation}
F_{N}\left(n<n_{c};\beta\right)=f_{n}\left(n,1;\beta\right)=-n\log\left[A\zeta(\beta C)\right],\label{eq_C:FN_lowT}
\end{equation}
which is a linear function of $n$. This linearity is again a manifestation
of the same coexistence between normal gas of microscopic domains
(with domains density $n=n_{c}$) and a single macroscopic domain
(with $n=\frac{1}{L}\rightarrow0$). The analysis is valid both above
and below $\beta_{c}$. For high temperatures ($\beta<\beta_{c}$)
the slope is negative (as $\zeta\left(\beta C\right)$ is a decreasing
function of $\beta$) and hence $n=0$ is disfavored. At the critical
temperature $A\left(\beta_{c}\right)\zeta\left(\beta_{c}C\right)=1$
by Eq.(\ref{eq_C:FM_betac}). Hence $F_{N}\left(n<n_{c}\right)=0$,
implying coexistence between the gas ($n=n_{c}$) and the condensate
($n=0$) in the unconstrained system. For $\beta>\beta_{c}$ the slope
is positive which implies that the thermodynamically stable phase is one with $n=0$.

For $\beta C\le2$ one has $n_{c}=0$.
From (\ref{eq_C:FN_saddlepoint}-\ref{eq_C:FN_highT}) it can be deduced
that $n^{*}$, the minimum of $F_{N}$, is a decreasing function of
temperature. If $c\equiv\beta_{c}C\le2$ it implies that $n\rightarrow0$
in a continuous manner as $\beta\rightarrow\beta_{c}$. On the other
hand, if $c>2$ then at high temperatures there is no linearity (no
coexistence with a condensate phase) but at temperatures $\beta=2/C<\beta_{c}$
a coexistence initiates for small $n$ or for large $m$.

In summary the free energy $F_{N}\left(n;\beta\right)$ is given by
Eqs.(\ref{eq_C:FN_fn_res}-\ref{eq_C:FN_highT}) for $n>n_{c}$, and
is a linear function given by Eq.(\ref{eq_C:FN_lowT}) for $n<n_{c}$,
where $n_{c}$ is given by Eq.(\ref{eq_C:nc_val}). For $c\le2$,
$n_{c}=0$ and hence there is no linear part. The function $W_{n}\left(n,\beta\right)$
appearing in Eq.(\ref{eq_res:zstar}) is defined by Eq.(\ref{eq_C:FN_saddlepoint})
for $n>n_{c}$ and by $z^{*}=1$ for $n\le n_{c}$.

\subsection{Finite size corrections\label{sub:Finite-size-corrections}}

The above analysis establishes the fact that $m$ is discontinuous at the transition
for all $c\equiv\beta_{c}C>1$. Moreover, while $n$ is discontinuous
for $c>2$, it is continuous for $c\le2$. The canonical analysis
presents a scenario which seems like a conventional first order transition.
But as discussed above, even though the order parameter is discontinuous,
at the transition there are features of a critical transition such
as divergence of correlation length and diverging susceptibility (for
$c<3$). How can this be explained in the framework of the canonical
analysis, i.e. as an outcome of the form of the ``Landau free energy''?
This question is answered by looking into the finite size corrections
to the large deviations functions, which turn up to be logarithmic
in $L$ as we shall show below. These logarithmic barriers can be
understood as the usual surface energy between coexisting phases that
appears in first order transitions. In $d$-dimensional models with short range interactions the surface energy scales like $L^{d-1}$, hence
% and which scales like $L^{d-1}$, and hence
at $d=1$ it is of order $O(1)$ or $O(\log(L))$. However
this is a somewhat simplistic argument since in this case the logarithmic scaling
relies also on the effective $r^{-2}$ interaction between spins.
Below we shall derive the finite size correction for $Z_{N}$ and
$Z_{M}$ and show that at and below the critical temperature there
are logarithmic barriers.

\subsubsection{Corrections to $Z_{N}$}

To go beyond the LDF result (\ref{eq_C:FN_highT}) when a saddle point
exists, i.e. for $n>n_{c}$, we note that the saddle point method
yields a sub-leading term due to the second derivative of the integrand,
i.e.
\[
Z_{N}\left(L,Ln;\beta\right)\approx\sqrt{\frac{2\pi}{L\left|\partial_{z}^{2}f_{n}\left(n,z^{*};\beta\right)\right|}}\frac{\left[A\Phi_{\beta C}\left(z^{*}\right)\right]^{Ln}}{Az^{*}{}^{L+1}}.
\]
Calculating explicitly $\partial_{z}^{2}f_{n}$ yields
\begin{equation}
Z_{N}\left(L,Ln;\beta\right)\approx\left[2\pi L\left|\frac{1}{z^{*2}n}-\frac{\Phi_{\beta C}''\left(z^{*}\right)}{\Phi_{\beta C}\left(z^{*}\right)}n\right|\right]^{-\frac{1}{2}}\frac{\left[A\Phi_{\beta C}\left(z^{*}\right)\right]^{Ln}}{Az^{*}{}^{L+1}}.\label{eq:finit_size_ZN1}
\end{equation}
This implies that the correction to the extensive free energy is logarithmic
as stated above. For $n<n_{c}$ the correction comes from the explicit
result for the branch-cut integration (see Eq.(\ref{eq:APPC_Z_N_branchcut_res})),
i.e.
\begin{equation}
Z_{N}\left(L,Ln;\beta\right)\approx\frac{b_{\phi}(n)\Gamma\left(\beta C\right)}{\pi L^{\beta C-1}\left(b_{\Lambda}(n)\right)^{\beta C-2}}\left[A\zeta\left(\beta C\right)\right]^{Ln},\label{eq:finite_size_ZN2}
\end{equation}
($\Lambda$ here is not the same as the one define in section \ref{sub:Condensate-phase--}) with
\begin{eqnarray*}
b_{\Lambda}(n) & = & 1-\frac{n\zeta\left(\beta C-1\right)}{\zeta\left(\beta C\right)},\\
b_{\phi}(n) & = & n\frac{\pi}{\Gamma\left(\beta C\right)\zeta\left(\beta C\right)}.
\end{eqnarray*}
For $n=n_{c}=\zeta\left(\beta C\right)/\zeta\left(\beta C-1\right)$
(Eq.(\ref{eq_C:nc_val})), $b_{\Lambda}\left(n_{c}\right)=0$ and
hence the approximation breaks down near $n=n_{c}$. The breakdown
of the branch cut integration approximation is due to the proximity
of the branch point and pole when $n\approx n_{c}$. Calculating the
finite size corrections for $n\approx n_{c}$ requires a different
approach, which will not be discussed here.

The free energy resulting from the approximation of Eqs.(\ref{eq:finit_size_ZN1}-\ref{eq:finite_size_ZN2})
with $L=1000$ is plotted in Fig.\ref{fig:LDF_finite}a, together
with an exact evaluation of the partition function for the same value
of $L$. The exact evaluation was done numerically as described in
Appendix D. Comparing the approximation and the exact evaluation it
can be seen that the approximations are quite accurate, except
around $n=n_{c}$. At the critical point, i.e. when
$z^{*}=1$ and $A\zeta\left(\beta_{c}C\right)=1$, $Z_{N}$ contain
only powers of $L$ and no exponential (or stretched exponential)
terms. This indicates that fluctuations of $n$ would be distributed
asymptotically as a power law and hence that the transition will be
scale invariant. Together with the discontinuity in $n$ (for $c>2$)
this implies that the transition is MOT.

To verify that the free energy barriers are logarithmic (corresponding
to power law dependence of $Z_{N}$ in $L$ ) we define $R$ as in
(\ref{eq_res:R}) to be the integral of $\tilde{F}_{N}=-\frac{1}{L}\log Z_{N}\left(L,N\right)$
from $n=0$ to the minimum of $\tilde{F}_{N}$ at $n^{*}$. From Eqs.(\ref{eq:finit_size_ZN1}-\ref{eq:finite_size_ZN2})
it can be seen that $\tilde{F}_{N}$ has a single minimum at $n^{*}>n_{c}$
and hence $R$ is well defined. We then evaluate $R$ numerically
and plot it in Fig.\ref{fig:finite_size_SN}. From this figure it
is apparent that $R$ scales logarithmically. This confirms that any
free energy barrier between $0$ and $n^{*}$ is logarithmic in $L$.

\subsubsection{Corrections to $Z_{M}$}

Following similar steps as in the previous section we can find the finite
size corrections for $Z_{M}$, but in this case the approximation
fails near the critical temperature and below it because Eq.(\ref{eq_C:FM_pole})
has no solution. Hence instead we shall use a simple argument to obtain
the leading order finite size behavior of $Z_{M}$ well below the
critical temperature. Specifically, in the regime where the gas phase
is unstable, we can expect that any magnetization is realized (to
leading order) by a phase separation of the chain to a $+$ condensate
and a $-$ condensate, their lengths are set by the condition that
the given magnetization is realized. This simply implies
\begin{equation}
Z_{M}\left(L,m;\beta>\beta_{c}\right)\sim\left(\frac{1+m}{2}L\right)^{-\beta C}\left(\frac{1-m}{2}L\right)^{-\beta C}.\label{eq:finit_size_ZM}
\end{equation}
This approximation is plotted in Fig.\ref{fig:LDF_finite}b along
with the exact numerical evaluation of the free energy (see Appendix
D) for $L=1000$, and seems to fit very well. Here again we see that
there are no exponential terms, and magnetization fluctuations are
suppressed only in a power law manner and not exponentially.

\section{\label{sec:RG-analysis}RG analysis}

As was shown above, many features of the model can be obtained analytically
in more than one way. However, the main tool used to study the IDSI
model and related models was the renormalization group (RG) analysis
\cite{anderson1970exact,cardy1981one}. To study the relation
of our model to those models we shall carry out an (approximate) RG
analysis to our model and compare it to the RG of the IDSI model,
stressing the similarities and differences.

\subsection{Model definition}

For the scaling --- or renormalization group (RG) --- analysis, it
is useful to define an off-lattice version of the model (\ref{eq_model:ttt_Hamiltonian2}).
This version also makes the connection to coulomb gas models and the
Kosterlitz-Thouless scenario more explicit.

We consider a gas of $N$ particles on an interval $\left[0,L\right]$,
or on the circle if periodic boundary conditions are considered. The
particles represent the domain boundaries (kinks) of the lattice version,
and following (\ref{eq_model:ttt_Hamiltonian2}), each pair of nearest
neighboring particles interact through an attractive logarithmic potential. To
avoid divergences there is an ultraviolet cutoff scale $a$ which
is the hard-core of the particles, so that the Hamiltonian takes the
form
\[
H\left(\left\{ r_{i}\right\} ;N\right)=C\sum_{i=1}^{N}\log\left(\frac{r_{i+1}-r_{i}}{a}\right)+N\Delta\qquad with\ \ \left|r_{i+1}-r_{i}\right|\ge a
\]
This constitutes the Coulomb gas picture of the model.

\subsection{RG analysis}

The grand canonical partition function takes the form
\begin{equation}
\fl Z_{C}\left(L;A,K\right)=\sum_{N=1}^{\infty}A^{N}\int_{-\infty}^{\infty}\frac{dr_{1}}{a}\int_{-\infty}^{\infty}\frac{dr_{2}}{a}...\int_{-\infty}^{\infty}\frac{dr_{N}}{a}\prod_{i=1}^{N}\left(\frac{r_{i+1}-r_{i}}{a}\right)^{-K}\Theta\left(r_{i+1}-r_{i}-a\right),\label{eq_rg:RG_ZC}
\end{equation}
where $A\equiv e^{-\beta\Delta}$ and $K=\beta C$. This integral
can be calculated exactly, as we have done above, but here we shall
follow the RG protocol of \cite{cardy1981one}: First we rescale the
core-size $a\rightarrow ae^{\kappa}$ for $0<\kappa\ll1$. Then we
express the partition sum in terms of rescaled parameters $A_{\kappa}$
and $K_{\kappa}$ such that in terms of these parameters it has the
same form as the original partition sum. This implies
\[
\fl Z_{C}\left(L;A,K\right)=\sum_{N=1}^{\infty}A_{\kappa}^{N}e^{N\kappa\left(K_{\kappa}-1\right)}\int_{-\infty}^{\infty}...\int_{-\infty}^{\infty}\prod_{i=1}^{N}\frac{dr_{i}}{a}\left(\frac{r_{i+1}-r_{i}}{a}\right)^{-K_{k}}\Theta\left(r_{i+1}-r_{i}-ae^{\kappa}\right).
\]
Applying the expansion
\[
\Theta\left(r_{i+1}-r_{i}-ae^{\kappa}\right)=\Theta\left(r_{i+1}-r_{i}-a\right)-a\kappa\delta\left(r_{i+1}-r_{i}-a\right)+O\left(\kappa^{2}\right),
\]
we get to first order in $\kappa$
\begin{eqnarray}
\fl Z_{C}\left(L;A,K\right) & = & Z_{C}\left(L;A_{\kappa}e^{\kappa\left(K_{k}-1\right)},K_{\kappa}\right)\nonumber \\
\fl  &  & -a\kappa\sum_{N=1}^{\infty}A_{\kappa}^{N}\sum_{j}\int_{-\infty}^{\infty}\prod_{i\neq j,j-1}\left[\frac{dr_{i}}{a}\left(\frac{r_{i+1}-r_{i}}{a}\right)^{-K_{k}}\Theta\left(r_{i+1}-r_{i}-a\right)\right]dr_{j-1}\times\nonumber \\
\fl  &  & \int_{r_{j-1}+a}^{r_{j+1}-a}\left(\frac{r_{j+1}-r_{j}}{a}\right)^{-K_{k}}\left(\frac{r_{j}-r_{j-1}}{a}\right)^{-K_{k}}dr_{j}\delta\left(r_{j+1}-r_{j}-a\right),\label{eq_rg:ZC_trans}
\end{eqnarray}
The integral in the last line of (\ref{eq_rg:ZC_trans}) results in
\[
\left(\frac{r_{j+1}-r_{j-1}-a}{a}\right)^{-K}\Theta\left(r_{j+1}-r_{j-1}-2a\right).
\]
Up to this point the calculation is exact, but here we should introduce
a physical argument which enables one to write closed RG equations. The
physical picture is of kinks interacting in nearest-neighbor pairs.
If the density of kinks is small (i.e. $A\ll a^{-1}$) it implies
that the typical distance between a pair is large and hence it is
plausible to assume $r_{j+1}-r_{j-1}-a\approx r_{j+1}-r_{j-1}$. Therefore
the result (to first order in $\kappa$) is
\begin{eqnarray*}
Z_{C}\left(L;A,K\right) & = & \sum_{N=1}^{\infty}A_{\kappa}^{N}\left(e^{N\kappa\left(K_{\kappa}-1\right)}-a\kappa NA_{\kappa}\right)\times\\
 &  & \qquad\int_{-\infty}^{\infty}...\int_{-\infty}^{\infty}\prod_{i=1}^{N}\frac{dr_{i}}{a}\left(\frac{r_{i+1}-r_{i}}{a}\right)^{-K_{k}}\Theta\left(r_{i+1}-r_{i}-a\right).
\end{eqnarray*}
To compensate for the additional factor $\left(e^{N\kappa\left(K_{\kappa}-1\right)}-a\kappa NA_{\kappa}\right)$
such that the partition sum retains its original form, the renormalized
parameters $A_{\kappa}$ and $K_{\kappa}$ are taken to be in leading
order in $\kappa$
\begin{eqnarray*}
A_{\kappa} & = & A\left(1+\kappa\left(1-K+aA\right)\right)+O\left(\kappa^{2}\right),\\
K_{\kappa} & = & K+O\left(\kappa^{2}\right).
\end{eqnarray*}
Defining $x\equiv1-K$ and $y\equiv aA$ the resulting flow equations
are
\begin{eqnarray}
\frac{dy}{d\kappa} & = & xy+y^{2},\label{eq_rg:RG_y}\\
\frac{dx}{d\kappa} & = & 0.\label{eq_rg:RG_x}
\end{eqnarray}
The physical intuition behind these flow equations is the following:
A change of scale has two effects. First, due to the change of scale
the density --- and hence the fugacity $y$ --- renormalizes. This
is accounted by the first term of (\ref{eq_rg:RG_y}). Second, rescaling
of the core size might cause near-by kinks to merge. The likelihood
of having two adjacent kinks scales as $y^{2}$, implying the second
term. For far away (but nearest neighbors) kinks, the effect of two
kinks that merged is the same as a single kink, as the interaction
is anyway between nearest neighbors. Since these
are the only outcomes of the scale transformation (to leading order),
the interaction $x$ is kept constant.

\begin{figure}
\includegraphics[scale=0.5]{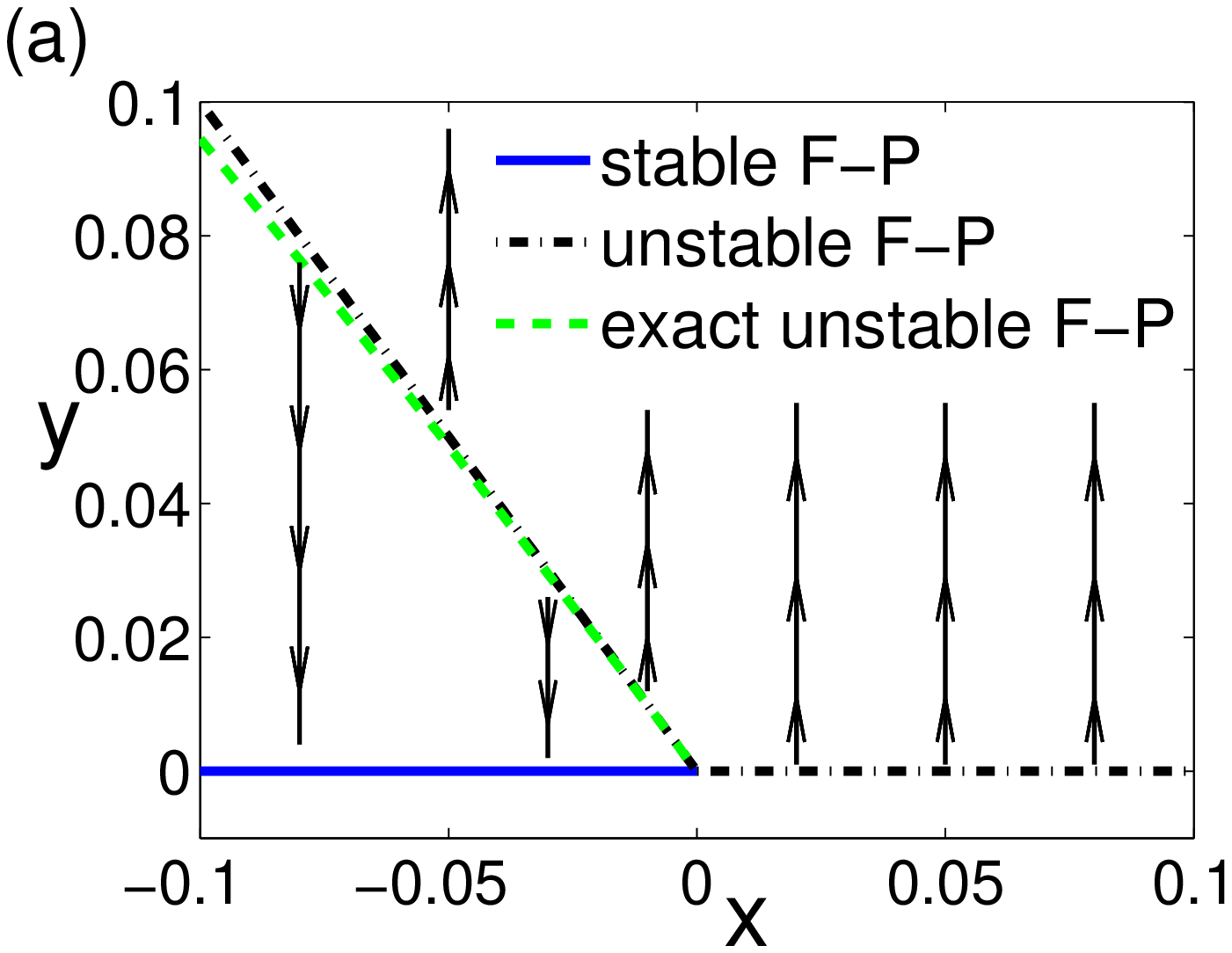}\includegraphics[scale=0.5]{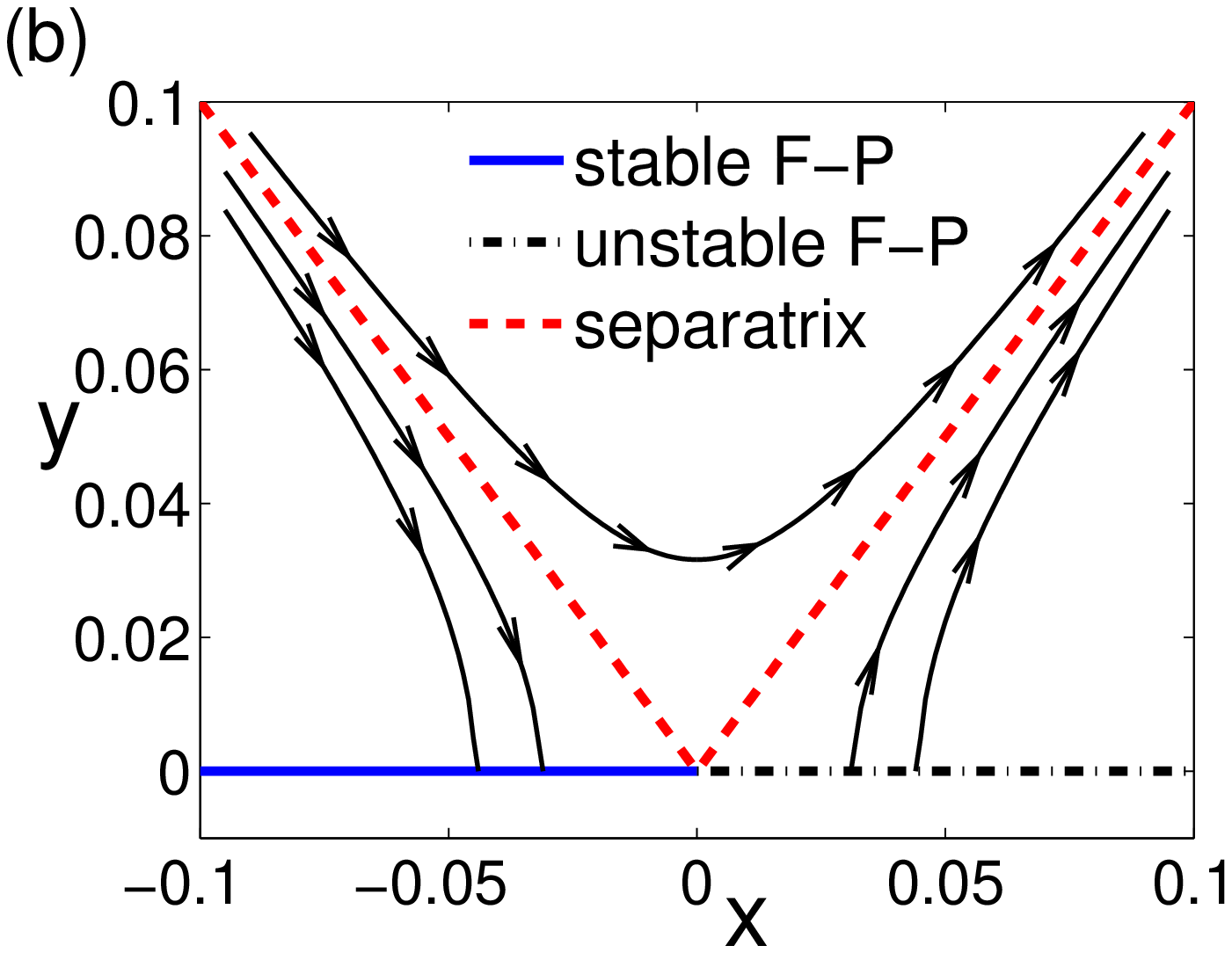}

\caption{\label{fig:RG-Flow}(color online) RG flow for (a) TIDSI model, Eqs.(\ref{eq_rg:RG_y}-\ref{eq_rg:RG_x})
and (b) IDSI model (or XY model), Eq.(\ref{eq_rg:IDSI_y}-\ref{eq_rg:IDSI_x}). }
\end{figure}

The RG flow diagram has a line of fixed points at $y=0$ which are
stable for $x<0$ and are unstable for $x>0$. It has another line
of unstable fixed points at $y=-x$ for $x<0$, which is the line
of phase transition: flow lines which start above this line increase
in $y$ until the validity of the analysis breaks down (the condition $y\ll1$ is no longer
valid) and hence flow into the disordered phase, while flow lines
which start below this line flow parallel to the $y$ axis into the
corresponding $y=0$ fixed point, which is the ordered phase. The
details of the flow diagram are presented in Fig.\ref{fig:RG-Flow}a
together with the exact phase transition line.

From the RG equations we can calculate the behavior of the correlation
length near the transition. This is done by integrating the flow equations,
starting just above the critical line $y^{*}=-x$, i.e. at $y=-x+\delta y$
with $\delta y\sim T-T_{c}$, and ending at $y=1$, where the analysis
breaks down but the correlations are order $1$. Then Eq.(\ref{eq_rg:RG_y})
yields
\begin{eqnarray*}
d\kappa & = & \frac{dy}{xy+y^{2}}\Rightarrow\\
\kappa & = & \left[\frac{1}{x}\log\left(\frac{y}{x+y}\right)\right]_{-x+\delta y}^{1}.
\end{eqnarray*}
Using $\xi\sim e^{\kappa}$ and $\delta y\sim T-T_{c}$, the leading
order term is
\[
\xi\sim\left(\frac{T-T_{c}}{\left|x\right|}\right)^{\frac{1}{x}}.
\]
Using $x=1-\beta_{c}C$, this is the same as Eq.(\ref{eq_res:xi}). The exponent
$1/x$ diverges for $x\rightarrow0_{-}$ suggesting that in this limit an essential
singularity develops.

\subsection{Comparison with the flow diagram of IDSI and related models}

The IDSI model, like other models with $r^{-2}$ interactions, has
a very similar representation as an interacting gas of charges \cite{cardy1981one}.
The only difference between TIDSI model and the IDSI model is that
in the IDSI all the charges interact with each other logarithmically,
while in the TIDSI model the interactions are only between nearest
neighbors. This implies that in the IDSI model, when two opposite
charges are close by they screen the effect of each other, and therefore
to leading order they cancel out and they contribute only a dipole
moment. This dipole moment turns out to renormalize the logarithmic
interaction between other charges, exactly in the same manner that
the screening of close by vortices renormalizes the interactions between
vortices in the XY model. Hence the $y^{2}$ term in (\ref{eq_rg:RG_y}),
which appears due to merging of two charges, is transferred to the
flow equation for $x$ and the resulting flow equations for the IDSI
model read
\begin{eqnarray}
\frac{dy}{d\kappa} & = & xy,\label{eq_rg:IDSI_y}\\
\frac{dx}{d\kappa} & = & y^{2},\label{eq_rg:IDSI_x}
\end{eqnarray}
where $x$ and $y$ are defined essentially the same as above. This
flow has only one line of fixed points at $y=0$, but it also has
a separatrix at $y=\left|x\right|$. The phase transition is along
the line $y=-x$ for $x<0$, where flow lines starting below it flow
to the $y=0$ fixed line (ordered phase) while flow lines that start
above it flow to the disordered phase, but the transition is controlled
by a single critical point $x=y=0$. The flow in the IDSI and TIDSI
case can be compared in Fig.\ref{fig:RG-Flow}. A consequence of the
different flow equations is the behavior of the correlation length,
which diverges algebraically in the TIDSI model while it exhibits
an essential singularity for the IDSI model. This is an outcome of
the flow near the critical point (or critical line in the case of
the TIDSI model): while linearizing Eq.(\ref{eq_rg:RG_y}-\ref{eq_rg:RG_x})
near $y=-x$ line we find that the flow is linear in the TIDSI model,
the flow is obviously quadratic in the IDSI case. This difference
yields the different behavior of the correlation length.

It is interesting to note that the lack of renormalization of the
coupling constant $x$ appears also in the context of discrete Gaussian
chain \cite{slurink1983roughening}
\[
H=-\sum_{ij}J_{ij}\left(h_{i}-h_{j}\right)^{2}\quad;\quad J_{ij}\sim\left|i-j\right|^{-2},
\]
where the height variables $h_{i}$ are integers, and with the boundary
condition $h_{0}=0$. This problem can also be mapped onto a dissipative
quantum particle in a periodic potential \cite{guinea1985diffusion}.
In this case there are infinitely many fugacities $y_{k}$ corresponding
to kinks with $h_{i}-h_{j}=k$, all renormalize due to both the density
scaling and merging of kinks, but the coupling coefficient does not
rescale. The connection between the TIDSI model and the discrete Gaussian
chain can be a subject of future investigations.

\section{\label{sec:Generalizations}Generalizations}

The TIDSI model can be generalized without loosing its solubility.
Below we first generalize the decay of interactions between spins
beyond the inverse square law of $J(r)$ in (\ref{eq_model:ttt_Hamiltonian1}).
We then consider the inverse squared law but with spin models other
than spin $\frac{1}{2}$ Ising model, namely Potts model and general
Ising model.

\subsection{\label{sub:General-interactions-decay}General interactions decay
law}

\subsubsection{Definition}

In this section we consider the Hamiltonian (\ref{eq_model:ttt_Hamiltonian1})
with
\begin{equation}
J\left(r\right)=Cr^{-\alpha},\label{eq_gen:E_LR_alpha}
\end{equation}
where $\alpha >1$. The long-range self energy of a domain $H_{LR}$ can be estimated
as before to be
\begin{equation}
H_{LR}\left(l\right)=-C\sum_{k=1}^{l}\frac{l-k}{k^{\alpha}}=-C\zeta\left(\alpha\right)l+C\left(\frac{1}{\alpha-1}+\frac{1}{\alpha-2}\right)l^{2-\alpha}+O\left(1\right).
\end{equation}
As before, the linear term will contribute a constant in the total energy. For $\alpha>2$ the subleading term is $O(1)$, and hence there is no transition.
%For $\alpha>2$ it can be seen that the energy is linear plus subleading
%corrections, and hence there is no transition.
We thus restrict the
discussion in this paper to $1<\alpha<2$. After readjusting the ground
state energy, the Hamiltonian reads
\begin{equation}
H^{\left(\alpha\right)}\left(\left\{ l_{a}\right\} ;N\right)=C_{\alpha}\sum_{a=1}^{N}l_{a}^{2-\alpha}+\Delta N,\label{eq_gen:h_alpha}
\end{equation}
with $C_{\alpha}\equiv C\left(\frac{1}{\alpha-1}+\frac{1}{\alpha-2}\right)$
and $\Delta=2J_{NN}$.

\subsubsection{Analysis}

\begin{figure}
\begin{centering}
\includegraphics[scale=0.5]{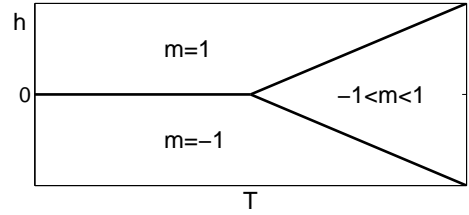}
\par\end{centering}

\caption{\label{fig:alpha-ising} Schematic phase diagram for model (\ref{eq_gen:h_alpha}).}
\end{figure}

The analysis of the above model in the grand canonical ensemble is
very similar to the one done in section \ref{sec:Grand-canonical-analysis}.
Skipping some details, the generating function is now
\begin{eqnarray}
Q\left(p,h,\mu;\beta\right) & = & \frac{e^{2\beta\left(\mu-\Delta\right)}W_{\alpha}\left(p+h\right)W_{\alpha}\left(p-h\right)}{1-e^{2\beta\left(\mu-\Delta\right)}W_{\alpha}\left(p+h\right)W_{\alpha}\left(p-h\right)},\label{eq_gen:alpha_Q}
\end{eqnarray}
with
\begin{equation}
W_{\alpha}\left(x\right)=\sum_{l}\exp\left(\beta xl-\beta C_{\alpha}l^{2-\alpha}\right)\equiv\Psi_{\beta C_{\alpha}}^{\alpha}\left(e^{\beta x}\right).\label{eq_gen:alpha_W}
\end{equation}
Note that the functions $W_{\alpha}$ and $\Psi_{\beta C}^{\alpha}$
have nothing to do with the functions $W_{\pm}\left(\right)$, $W_{n}\left(\right)$
and $\Psi_{\pm}\left(\right)$ defined in previous sections. The relevant properties
of $\Psi_{\gamma}^{\alpha}\left(u\right)$ for $1<\alpha<2$ are:
\begin{enumerate}
\item $\Psi_{\gamma}^{\alpha}\left(u\right)$ is analytic in the complex
plane except for a branch-cut along $[1,\infty)$
\item $\frac{d^{n}}{du^{n}}\left.\Psi_{\gamma}^{\alpha}\left(u\right)\right|_{u=1}<\infty$
for any $n\ge0$.
\end{enumerate}
Due to the similarity between the properties of $\Psi_{\gamma}^{\alpha}(u)$ and those of the polylogarithm function $\Phi_{\gamma}(u)$, the analysis of this case is very similar to the one done for the $\alpha=2$ case.
Hence the analysis done above for the case $\alpha=2$ holds also
for this case. The critical temperature for $\mu=0$ and $h\ge0$
is given by
\[
\Psi_{\beta C_{\alpha}}^{\alpha}\left(1\right)\Psi_{\beta C_{\alpha}}^{\alpha}\left(e^{\beta(p-h)}\right)=e^{2\beta\Delta},
\]
and specifically for $h=0$ it is
\[
e^{\beta\Delta}=\Psi_{\beta C_{\alpha}}^{\alpha}\left(1\right)\equiv\sum_{l}\exp\left(-\beta C_{\alpha}l^{2-\alpha}\right).
\]
Due to the finiteness of the derivatives of $\Psi_{\gamma}^{\alpha}$,
the transition is discontinuous (both in $m$ and $n$) for any value
of $C_{\alpha}$ and $h$. A schematic phase diagram is depicted in
Fig.\ref{fig:alpha-ising}. The $h=0$ transition line between the two magnetically saturated phases is an ordinary
first order phase transition, while the two transition lines separating the magnetically unsaturated phase from the saturated ones are MOT, like in the $\alpha=2$ case. However, unlike the latter
case, the distribution of domain sizes is not power law even at the
transition, and instead it takes a form similar to Eq.(\ref{eq_GC:domain_size_dist})
\begin{eqnarray}
P_{\pm}\left(l\right) & \simeq & e^{\beta\left(p^{*}\pm h\right)l}e^{-\beta C_{\alpha}l^{2-\alpha}}\equiv\exp\left[-\frac{l}{\xi_{\pm}}-\left(\frac{l}{\xi_{\alpha}}\right)^{2-\alpha}\right].\label{eq_gen:domain_size_dist}
\end{eqnarray}
The exponential length scales $\xi_{\pm}$ diverges at the transition
just like in the $\alpha=2$ case and hence the transition is MOT.
However, at the transition $P_{\pm}(l)$ takes a stretched exponential form
and hence all its moments are finite, unlike in that case. The stretched
exponential law defines a different length scale $\xi_{\alpha}\equiv\left(\beta C_{\alpha}\right)^{\frac{1}{2-\alpha}}$,
which sets the scale of correlations. It can be seen that indeed this
length scale diverges for $\alpha=2$ and $\beta C_{\alpha}>1$.

\subsection{\label{sub:General-spins}General spins}

% Preview source code from paragraph 127 to 136

We return now to $J(r)\approx C r^{-2}$ case, but consider more general
spin models. For concreteness, we focus on Potts spins and general
Ising spins, but other models can be analyzed following the same steps.
Specifically, we show that such models can be solved exactly using
the transfer matrix approach.

\subsubsection{Potts spins}

We consider now a chain of $L$ spins $\sigma_{i}$ that
can take a value in $\left[1..K\right]$ with a Hamiltonian analogous
to Eq.(\ref{eq_model:ttt_Hamiltonian1}), i.e.

\begin{equation}
H=-J_{NN}\sum\delta_{\sigma_{i},\sigma_{i+1}}-\sum_{i<j}J(i-j)\delta_{\sigma_{i}\sigma_{j}}I\left(i\sim j\right).\label{eq_gen:ttt_Hamiltonian1}
\end{equation}
Due to the truncation of the LR interactions the term $\delta_{\sigma_{i}\sigma_{j}}$ within each domain
is always unity. The model can be casted in the domains representation,
in which a configuration is composed of $N$ domains with sizes $\left\{ l_{n}\right\} $
and spins $\left\{ s_{n}\right\} $. A domain self energy then has
a very similar form to Eqs.(\ref{eq_model:H_NN},\ref{eq_model:H_LR}),
which yields the same Hamiltonian as (\ref{eq_model:ttt_Hamiltonian2}),
i.e.
\begin{equation}
H\left(\left\{ l_{n}\right\} ,\left\{ s_{n}\right\} ;N\right)=C\sum_{n=1}^{N}\log\left(l_{n}\right)+\Delta N+Const.\label{eq_gen:ttt_Hamiltonian2}
\end{equation}

As there are $K$ symmetric spin states in this case and not only
two, a natural ensemble to consider is one
where the set $\left\{ L_{s}\right\} _{s=1}^{K}$ is fixed, where
$L_{s}$ is the total number of spins of type $s$. The partition
function of such an ensemble, where in addition $N$ is fixed, is
\[
\fl Z_{0}\left(\left\{ L_{s}\right\} ,N;\beta\right)=\sum_{N}\sum_{\left\{ l_{n}\right\} }\sum_{\left\{ s_{n}\right\} }\prod_{n=1}^{N}\frac{e^{-\beta\Delta}}{l_{n}^{\beta C}}I\left(s_{n}\neq s_{n-1}\right)\prod_{s=1}^{K}I\left(\sum l_{n}\delta_{s_{n},s}=L_{s}\right).
\]
A set of $K$ order parameters can be constructed as
\begin{equation}
m_{s}=\frac{KL_{s}-L}{L(K-1)}.\label{eq_gen:magnetization-1}
\end{equation}
To define the corresponding grand partition function we should therefore
introduce $K+1$ fugacities ($\mu$ and $p_{s}$, $s=1..K$) corresponding to
the $K+1$ constraints (fixed $N$ and $\left\{ L_{s}\right\} $).
Then the grand partition function is

\begin{eqnarray}
Q\left(p,h,\mu;\beta\right)  =  \sum_{N}\prod_{n=1}^{N}e^{\beta\left(\mu-\Delta\right)}\sum_{s_{n}\neq s_{n-1}}U_{\beta C}\left(p_{s_{n}}\right), \\
U_{\rho}\left(x\right)  \equiv  \sum_{l}\frac{e^{\beta xl}}{l^{\rho}}. \label{eq_gen:UbetaC}
\end{eqnarray}
To proceed we define the transfer matrix,

\begin{equation}
\hat{T}_{\sigma\tau}=\cases{
e^{\beta\left(\mu-\Delta\right)}U_{\beta C}\left(\beta p_{\tau}\right) & $\sigma\neq\tau$ \\
0 & $\sigma=\tau$
}.
\end{equation}
Assuming for simplicity fixed boundary conditions $s_{1}$ and
$s_{N}$, and using conventional bra-ket notations, the grand-partition function can be expressed as
\begin{eqnarray}
Q\left(\left\{ p_{s}\right\} ,\mu;\beta\right) & = & \sum_{N=1}^{\infty}\left\langle s_{1}\right|\hat{T}^{N}\left|s_{N}\right\rangle \nonumber \\
 & = & \left\langle s_{1}\right|\frac{\hat{T}}{1-\hat{T}}\left|s_{N}\right\rangle .\label{eq_gen:Potts_GCPF}
\end{eqnarray}
We can first set $\mu=0$ and $p_{s}=p$ for all $s$, i.e. constraining
only the total size $L$. As before, the most negative singularity
of $Q$ in $p$ can stem either from the denominator, where the maximal
eigenvalue of the matrix $\hat{T}\left(p\right)$ satisfies $\lambda_{max}=1$,
i.e.
\begin{equation}
p^{*}=\arg\min_{p}\left\{ \lambda_{max}(p)=1\right\} ,\label{eq:gs_p_highT}
\end{equation}
or from the branch point of $U_{\beta C}$, at
\begin{equation}
p^{*}=0\label{eq:gs_p_lowT}.
\end{equation}
In this setting the eignevalues of $\hat{T}$ can be obtained exactly,
and the maximal eigenvalue is
\[
\lambda_{max}=\left(K-1\right)e^{-\beta\Delta}U_{\beta C}\left(p\right).
\]
Hence for low enough temperature, $p^{*}=0$, which as in the TIDSI
implies a condensation transition.

To see that indeed there is condensation,
we can set $p_{1}=p+r$ and $p_{s}=p$ for all $s>1$. Then
\[
\frac{L_{1}}{L}=-\left.\frac{dp^{*}}{dr}\right|_{r=0}.
\]
The largest eigenvalue of $\hat{T}$ is given by
\[
\fl \lambda_{max}=\frac{e^{-\beta\Delta}}{2}\left[\left(K-2\right)U_{\beta C}\left(p\right)+\sqrt{\left(K-2\right)^{2}U_{\beta C}\left(p\right)^{2}+\left(4K-4\right)U_{\beta C}\left(p\right)U_{\beta C}\left(p+r\right)}\right].
\]
Hence in the high temperature phase, for which $p^{*}$ is set by
the condition $\lambda_{max}=1$, we find by straightforward calculation
\[
L_{1}=\frac{L}{K}\quad\Rightarrow\quad m_{1}=0,
\]
while in the low temperature phase $p^{*}=-r$ (for $r>0$) and therefore
\[
L_{1}=L\quad\Rightarrow\quad m_{1}=1.
\]
This implies that the order parameter $m_{1}$ jumps from $0$ to 1 at the transition
just like in the original TIDSI. Finally the domain size distribution
(at $r=0$) can be written, in analogy with Eq.(\ref{eq_GC:domain_size_dist}), as

\[
P\left(l\right)\simeq\frac{Z_{C}\left(L-l,h\right)}{Z_{C}\left(L,h\right)}\times\frac{1}{l^{\beta C}}=\frac{e^{-\beta p^{*}l}}{l^{\beta C}}=\frac{e^{-l/\xi}}{l^{\beta C}},
\]
with $\xi=\log\left(p^{*}\right)^{-1}$ being the diverging length
scale. Hence the transition is MOT of the same kind as in the original
TIDSI.

\subsubsection{General Ising spins}

Now we consider a spin $\frac{K}{2}$ Ising model, i.e. where each
spin $\sigma_i$ can take one the of $K+1$ different values $\{-K,-K+2,...,K\}$.
The Hamiltonian is

\begin{equation}
H=-J_{NN}\sum\sigma_{i}\sigma_{i+1}-\sum_{i<j}J(i-j)\sigma_{i}\sigma_{j}I\left(i\sim j\right).\label{eq_gen:ttt_Hamiltonian_Ising}
\end{equation}
Now, different domains have different energies according to their
spin. The Hamiltonian in the domains variables read
\[
\fl H\left(\left\{ l_{n}\right\} ,\left\{ s_{n}\right\} ;N\right)=-\sum_{n=1}^{N}B_1 s_{n}^{2}l_{n}+C\sum_{n=1}^{N}s_{n}^{2}\log\left(l_{n}\right)+\sum_{n=1}^{N}\left(B_2 s_{n}^{2}-J_{NN}s_{n}s_{n+1}\right)+Const,
\]
where $B_1$ and $B_2$ are positive coefficients. The grand canonical partition function has the
same form as in the Potts model, i.e. Eq.(\ref{eq_gen:Potts_GCPF}),
with the transfer matrix
\[
\hat{T}_{\sigma\tau}=\exp\left[\beta\left(\mu-B_2\tau^{2}+J_{NN}\sigma\tau\right)\right]U_{\beta C\tau^{2}}\left(p_{\tau}+B_1\tau^{2}\right),
\]
for $\sigma\neq\tau$ and $\hat{T}_{\sigma\sigma}=0$. The function $U_{\rho}(x)$ is defined by Eq.(\ref{eq_gen:UbetaC}). Setting $p_{\tau}=p$ and $\mu=0$, the thermodynamic limit is obtained
when either the maximal eigenvalue is unity, i.e.
\begin{equation}
\lambda_{max}\left(p^{*}\right)=1,\label{eq_gen:Ising_HT_cond}
\end{equation}
or at
\begin{equation}
p^{*}=-\max_{\tau}\left\{ B_1\tau^{2}\right\} =-B_1K^{2}.\label{eq_gen:Ising_LT_cond}
\end{equation}
Calculating $\lambda_{max}$ is hard and does not provide any new insights.
However as $\hat{T}$ is a non-negative irreducible matrix, the Perron-Furbenius
theorem implies that $\lambda_{max}>0$. Moreover, for $J_{NN}<B/K$
increasing $\beta$ decreases (or do not change) all the elements
of $\hat{T}$, thus by Wielandt's theorem \cite{meyer2000matrix}
$\lambda_{max}$ also decreases with $\beta$. The reverse goes for
increasing $p$, and hence in the high temperature phase, where $p^{*}$
is set by Eq.(\ref{eq_gen:Ising_HT_cond}), $p^{*}$ is an increasing
function of $\beta$. Therefore there is a critical $\beta$ for which
$\lambda_{max}\left(B_1K^{2}\right)=1$, which sets the transition
temperature. Adding to the Hamiltonian a magnetic field $h$ which
is coupled to the magnetization order parameter $m=\frac{1}{L}\sum l_{n}s_{n}$
amounts to setting $p_{\tau}=p+\tau h$. The maximal eigenvalue $\lambda_{max}$
must be symmetric with respect to $h\rightarrow-h$, and hence in the
high temperature regime $m=0$. In the low temperature regime $p^{*}=-B_1K^{2}-\left|h\right|K$
which implies that $m=\pm K$ as expected. Hence, while there is full
magnetization also in this model, the spin of the macroscopic domain is only
two fold degenerate, and not $K$ (or $K+1$) fold degenerate as in
the Potts model case.

\section{\label{sec:Conclusions}Conclusions}

In this paper we present a detailed analysis of the TIDSI model which
was recently introduced in \cite{bar2014mixed}. The study is motivated
by the observation that this is an exactly soluble model which exhibits
a mixed order transition and which serves as a link between different
classes of models exhibiting MOT. The steady state of the model and
its phase diagram are first calculated in the grand canonical ensemble.
In addition a canonical analysis which sheds new light on the mechanism
of a mixed order transition is presented. This analysis shows that
for $c>2$, where both order parameters, the magnetization $m$ and
the domains density $n$, are discontinuous, criticality stems from
logarithmic barriers in the effective Landau free energy. For $c<2$
where $n$ is continuous, the magnetization $m$ remains discontinuous
due to spin inversion symmetry in the high temperature phase. We also
elaborate on the RG analysis presented in \cite{bar2014mixed}, and
finally generalize the model by introducing general power-law decaying
interactions ($1/r^{\alpha}$) and several other types of spin variables.
These generalizations elucidate the special features of the borderline
case $\alpha=2$ and show that MOT can take place in a rather general class
of discrete spin models.

The TIDSI model provides a bridge between one dimensional models with
$1/r^{2}$ interactions such as the IDSI and one dimensional models
exhibiting the depinning transition, like the PS model. This opens
a window for a more general question regarding the connection between
models exhibiting mixed order transitions. For instance, the spiral
model of \cite{toninelli2006jamming,toninelli2007toninelli} is a
two dimensional model which exhibits MOT. Can the mechanism which
leads to the transition to the jammed state be related to that of
the one dimensional models exhibiting MOT discussed in this paper?
In the context of networks, there has been a recent debate \cite{achlioptas2009explosive,da2010explosive}
regarding the nature of transition of a process dubbed ``explosive
percolation'' in which an irreversible network evolution models exhibit
a rather abrupt appearance of a giant component. There is some evidence
\cite{bizhani2012discontinuous,tian2012nature} that this process,
or some version of it, leads to a mixed order transition, with a finite
size behavior similar to TIDSI (i.e. logarithmic barriers). Can this
process, and related percolation models like $k$-core percolation
\cite{liu2012core} be connected with our model? A general framework
for studying such mixed order transitions is still missing.

Another interesting and not thoroughly explored direction of research
has to do with the dynamics of (equilibrium) models exhibiting MOT.
Phase separation kinetics is the dynamical behavior of systems quenched
from a high temperature phase, usually infinite temperature, to a
low temperature ordered phase. The phase ordering kinetics of systems
exhibiting second order phase transitions, usually at zero temperature
\cite{bray1994theory} has been a subject of a large body of work
in recent years. However, it seems that there are no elaborate studies
of the phase ordering kinetics in models exhibiting MOT. In \cite{lee1993phase}
the phase ordering kinetics of the IDSI and other models with long-range
interactions were studied, but only at zero temperature, while a more
interesting case would be quenching to the critical temperature, in
which unlike in second order transitions, real phase ordering is expected.
Another intriguing question in this context is the connection of the
dynamics of the TIDSI model with nonequilibrium models with absorbing
states. This is left for future work.

We thank M. Aizenman, O. Cohen and O. Hirschberg for helpful discussions. The support of the Israel Science Foundation (ISF) and of the Minerva
Foundation with funding from the Federal German Ministry for Education and Research is gratefully acknowledged. We also thank the Galileo Galilei Institute for Theoretical Physics
for the hospitality and the INFN for partial support during the completion of this work.

\begin{appendices}

\section{Analysis of the LHS of Eq.(\ref{eq_C:FM_saddlepoint})}

We wish to show that
\[
g\left(z_{+}\right)\equiv\frac{\Phi_{\beta C-1}\left(z_{+}\right)\Phi_{\beta C}\left(z_{-}^{*}\left(z_{+}\right)\right)}{\Phi_{\beta C}\left(z_{+}\right)\Phi_{\beta C-1}\left(z_{-}^{*}\left(z_{+}\right)\right)},
\]
is an increasing function of $z_{+}$ for fixed $\beta$. The function
$z_{-}^{*}$ is given by the implicit relation (\ref{eq_C:FM_pole}),
i.e.
\[
\Phi_{\beta C}\left(z_{-}^{*}\right)=\left[A^{2}\Phi_{\beta C}\left(z_{+}\right)\right]^{-1},
\]
implying that $z_{-}^{*}$ is a decreasing function of $z_{+}$. The
function $g$ can be written as $g\left(z_{+}\right)=h\left(z_{+}\right)/h\left(z_{-}^{*}\right)$
where
\[
h\left(u\right)\equiv\frac{\Phi_{\beta C-1}\left(u\right)}{\Phi_{\beta C}\left(u\right)}.
\]
Showing that $h(u)$ is an increasing function thus proves that $g\left(z_{+}\right)$
is also an increasing function.

To show that $h$ is increasing we inspect its derivative
\[
\frac{dh(u)}{du}=\frac{\Phi_{\beta C-2}\left(u\right)\Phi_{\beta C}\left(u\right)-\Phi_{\beta C-1}\left(u\right)^{2}}{u\Phi_{\beta C}\left(u\right)^{2}}\equiv\frac{\mathcal{N}}{\mathcal{D}}.
\]
The denominator $\mathcal{D}$ is trivially positive and it is left
to show the same for the numerator $\mathcal{N}$:
\begin{eqnarray*}
\mathcal{N} & = & \sum_{k,l=1}^{\infty}\left[\frac{u^{l}}{l^{c-2}}\frac{u^{k}}{k^{c}}-\frac{u^{l}}{l^{c-1}}\frac{u^{k}}{k^{c-1}}\right]\\
 & = & \sum_{k,l=1}^{\infty}\frac{u^{l}}{l^{c-1}}\frac{u^{k}}{k^{c}}\left[l-k\right]\\
 & = & \sum_{k<l}^{\infty}u^{l+k}\left[l-k\right]\left[\frac{1}{l^{c-1}}\frac{1}{k^{c}}-\frac{1}{k^{c-1}}\frac{1}{l^{c}}\right]\\
 & = & \sum_{k<l}^{\infty}\frac{u^{l+k}}{l^{c}k^{c}}\left[l-k\right]^{2}>0.
\end{eqnarray*}
Q.E.D.

\section{Approximating (\ref{eq_C:FM_fmz_def}) without saddle point }

Here we calculate the integral (\ref{eq_C:FM_fmz_def}) when $f_{m}\left(m,z_{+}\right)$
has no saddle point for $\left|z_{+}\right|<1$, i.e. in the regime
$\beta<\beta_{c}$ and $m>m_{c}$, which implies $\beta C>2$. In
this case the contour of the integral can be deformed as presented
in Fig.\ref{fig:Countors}b. The contour $C_{bc}$ can be expressed
as concatenation of four parts which are
\begin{eqnarray}
(I) & : & \left[R-i\epsilon,1-i\epsilon\right],\nonumber \\
(II) & : & \left\{ 1-\epsilon e^{i\theta}:\frac{\pi}{2}<\theta<\frac{3\pi}{2}\right\}, \nonumber \\
(III) & : & \left[1+i\epsilon,R+i\epsilon\right],\nonumber \\
(IV) & : & \left\{ Re^{i\theta}:\delta<\theta<2\pi-\delta\right\} .\label{eq:contour_def}
\end{eqnarray}
where $R\gg1$ and $\epsilon\ll1$ are free parameters, and $tg\left(\delta\right)=\frac{\epsilon}{R}$.
Along this contour the function $f_{m}$ can have complex values,
and hence we define
\begin{equation}
\Lambda\equiv Re\left[f_{m}\right]\quad;\quad\phi\equiv Im\left[f_{m}\right].\label{eq:APPB_Lam_Phi_def}
\end{equation}
The integral can thus be written as
\begin{equation}
Z_{M}(L,M;\beta)=\frac{1}{2\pi iA^{3}}\oint_{C_{bc}}dz_{+}e^{-L\left[\Lambda\left(m,z_{+};\beta\right)+i\phi\left(m,z_{+};\beta\right)\right]}.\label{eq:APPB_ZM_def}
\end{equation}
The contribution of part $\left(II\right)$ of $C_{bc}$ is $O\left(\epsilon\right)$
and hence can be neglected in the limit $\epsilon\rightarrow0$. To
show the same for part $\left(IV\right)$ we note that for $\left|u\right|\gg1$,
$\left|\Phi_{\gamma}\left(u\right)\right|\sim\log\left(u\right)^{\gamma}$. Hence from (\ref{eq_C:FM_pole}) we see that for $\left|z_{+}\right|=R\gg1$,
$z_{-}^{*}\sim\left(\log R\right)^{-\gamma}$. Hence along part $\left(IV\right)$
the absolute value of the integrand scales as $\left(R^{-L_{+}}\log\left(R\right)^{\gamma L_{-}}\right)$
which vanishes faster than $R^{-2}$ for any extensive $L_{+}$ and
therefore the integral over $\left(IV\right)$ is zero. Hence only
parts $(I)$ and $(III)$ contribute.

The polylogarithm function, and hence also $f_{m}$, has a series
expansion with real coefficients. Therefore $f_{m}\left(m,\bar{z};\beta\right)=\overline{f_{m}\left(m,z;\beta\right)}$
where $\bar{z}$ is the complex conjugate of $z$. Parts $\left(I\right)$
and $\left(III\right)$ traverse complex conjugated paths (in reverse
order) and hence
\begin{equation}
Z_{M}(L,M;\beta)=\frac{1}{\pi A^{3}}\int_{1+i0}^{\infty+i0}dz_{+}e^{-L\Lambda}\sin\left(L\phi\right).\label{eq:APPB_ZM_I_and_III}
\end{equation}
This integral can be handled by a version of the stationary phase
approximation: If $\frac{d}{dz}\phi\left(m,z;\beta\right)\neq0$,
the rapid oscillations of the $\sin()$ function (due to the large
$L$) would imply that the neighborhood of $z$ has no contribution
to the integral. For $z=1$, $\frac{d}{dz}\phi\left(m,z;\beta\right)=0$,
and due to the factor $z_{+}^{-L_{+}}$ it will be the only extremum
contributing. Hence the integral can be limited to a small neighborhood
$\left(1,\eta\right)$, i.e.
\begin{equation}
Z_{M}(L,M;\beta)\approx\frac{1}{\pi A^{3}}\int_{1}^{1+\eta}dz_{+}e^{-L\Lambda}\sin\left(L\phi\right),\label{eq:APPB_ZM_eta}
\end{equation}
where $\eta$ will be set shortly. It is shown below (section B.1)
that
\begin{eqnarray}
\Lambda\left(m,1+\delta z;\beta\right) & \approx & \Lambda\left(m,1;\beta\right)+b_{\Lambda}\left(m;\beta\right)\delta z,\label{eq:APPB_Lam_expansion}\\
\phi\left(m,1+\delta z;\beta\right) & \approx & b_{\phi}\left(m;\beta\right)\delta z^{\beta C-1}.\label{eq:APPB_Phi_expansion}
\end{eqnarray}
Inserting (\ref{eq:APPB_Lam_expansion}-\ref{eq:APPB_Phi_expansion})
into (\ref{eq:APPB_ZM_eta}) yields
\begin{eqnarray*}
Z_{M}(L,M;\beta) & \approx & \frac{e^{-L\Lambda\left(m,1;\beta\right)}}{\pi A^{3}}\int_{0}^{\eta}d\delta ze^{-Lb_{\Lambda}\delta z}\sin\left(Lb_{\phi}\delta z^{\beta C-1}\right)\\
 & = & \frac{e^{-L\Lambda\left(m,1;\beta\right)}}{\pi A^{3}Lb_{\Lambda}}\int_{0}^{b_{\Lambda}L\eta}due^{-u}\sin\left(\frac{Lb_{\phi}}{\left(b_{\lambda}L\right)^{\beta C-1}}u^{\beta C-1}\right).
\end{eqnarray*}
As there is no saddle point for $z_{+}<1$, and $f_{m}(m,z\rightarrow0;\beta)\rightarrow-\infty$,
we see that $b_{\Lambda}>0$. In addition, Eq.(\ref{eq:APPB_Phi_expansion})
together with the condition $L\phi\left(m,1+\eta;\beta\right)=1$
implies $\eta\sim L^{\frac{1}{1-\beta C}}$. In the thermodynamic
limit the upper limit tends to infinity as $\beta C>2$. In addition,
due to $e^{-u}$ factor only $u=O(1)$ contributes, and in this region
the argument of the $\sin\left(\right)$ tends to 0 (as $\beta C>2$),
hence it can be expanded:
\begin{eqnarray*}
Z_{M}(L,M;\beta) & \approx & \frac{e^{-L\Lambda\left(m,1;\beta\right)}b_{\phi}}{\pi A^{3}L^{\beta C-1}b_{\Lambda}^{\beta C-2}}\int_{0}^{\infty}due^{-u}u^{\beta C-1}\\
 & = & \frac{b_{\phi}\Gamma\left(\beta C\right)}{\pi A^{3}L^{\beta C-1}b_{\Lambda}^{\beta C-2}}e^{-Lf_{m}\left(m,1;\beta\right)}.
\end{eqnarray*}
Q.E.D

\subsection{Deriving Eqs.(\ref{eq:APPB_Lam_expansion}-\ref{eq:APPB_Phi_expansion})}

We wish to prove (\ref{eq:APPB_Lam_expansion}-\ref{eq:APPB_Phi_expansion}).
We define $z_{-}^{*}(1+\delta)=z_{-}^{*}(1)+\chi$. Then Eq.(\ref{eq_C:FM_pole})
implies
\[
\Phi_{\beta c}\left(z_{-}^{*}(1)+\chi\right)=\frac{1}{A^{2}\Phi_{\beta c}(1+\delta)}.
\]
Expanding both sides in terms of $\chi$ and $\delta$ for $\beta C>2$
yields
\begin{eqnarray*}
\Phi_{\beta c}\left(z_{-}^{*}(1)+\chi\right) & \approx & \Phi_{\beta c}\left(z_{-}^{*}(1)\right)+\frac{1}{z_{-}^{*}}\Phi_{\beta c-1}\left(z_{-}^{*}(1)\right)\chi,\\
\frac{1}{\Phi_{\beta c}(1+\delta)} & \approx & \frac{1}{\zeta_{\beta c}+\zeta_{\beta c-1}\delta+i\pi\delta^{\beta c-1}/\Gamma(\beta c)}\\
 & \approx & \frac{1}{\zeta_{\beta c}}-\frac{\zeta_{\beta c-1}}{\zeta_{\beta c}^{2}}\delta-i\frac{\pi}{\zeta_{\beta c}^{2}\Gamma(\beta c)}\delta^{\beta c-1}.
\end{eqnarray*}
Hence
\begin{eqnarray*}
Re\left[\chi\right] & = & -\frac{\zeta_{\beta c-1}z_{-}^{*}(1)}{A^{2}\zeta_{\beta c}^{2}\Phi_{\beta c-1}\left(z_{-}^{*}(1)\right)}\delta + o(\delta),\\
Im\left[\chi\right] & = & -\frac{\pi z_{-}^{*}(1)}{\Phi_{\beta c-1}\left(z_{-}^{*}(1)\right)A^{2}\zeta_{\beta c}^{2}\Gamma(\beta c)}\delta^{\beta c-1} + o\left(\delta^{\beta c-1}\right).
\end{eqnarray*}
Inserting these results into the definition of $f_{m}$, i.e. Eq.(\ref{eq_C:FM_fmz_def}),
yields
\begin{eqnarray*}
f_{m}(m,1+\delta;\beta) & = & \frac{1+m}{2}\log\left(1+\delta\right)+\frac{1-m}{2}\log\left(z_{-}^{*}+\chi\right)\\
 & \approx & \frac{1-m}{2}\log\left(z_{-}^{*}\right)+b_{\Lambda}\left(m;\beta\right)\delta-ib_{\phi}\left(m;\beta\right)\delta^{\beta c-1},\\
b_{\Lambda}\left(m;\beta\right) & = & \frac{1+m}{2}-\frac{(1-m)\zeta_{\beta c-1}\Phi_{\beta c}\left(z_{-}^{*}\right)}{2\zeta_{\beta c}\Phi_{\beta c-1}\left(z_{-}^{*}\right)},\\
b_{\phi}\left(m;\beta\right) & = & \frac{\pi(1-m)}{\Phi_{\beta c-1}\left(z_{-}^{*}\right)A^{2}\zeta_{\beta c}^{2}\Gamma(\beta c)},
\end{eqnarray*}
with $z_{-}^{*}=z_{-}^{*}\left(1\right)$. Hence

\begin{eqnarray*}
\Lambda(m,1+\delta) & \approx & \Lambda(1)+b_{\Lambda}\left(m;\beta\right)\delta,\\
\phi(m,1+\delta) & \approx & b_{\phi}\left(m;\beta\right)\delta^{\beta c-1}.
\end{eqnarray*}
Q.E.D

\section{Approximating (\ref{eq_C:FN_fn_def}) without saddle point}

Here we calculate the integral (\ref{eq_C:FN_fn_def}) when $f_{n}\left(n,z\right)$
has no saddle point for $\left|z\right|<1$, i.e. in the regime $n<n_{c}$,
which implies $\beta C>2$. This case is very similar to the case
considered in Appendix B, and hence we skip some of the details. The
contour of the integral can deformed to a contour $C_{bc}$, defined
by Eq.(\ref{eq:contour_def}) and presented in Fig.\ref{fig:Countors}b.
Along this contour the function $f_{n}$ can have complex values,
and hence we define
\[
\Lambda\equiv Re\left[f_{n}\right]\quad;\quad\phi\equiv Im\left[f_{n}\right].
\]
The integral then can be written in a form equivalent to (\ref{eq:APPB_ZM_def}-\ref{eq:APPB_ZM_I_and_III})
\begin{eqnarray*}
Z_{N}(L,Ln;\beta) & = & \frac{1}{2\pi i}\oint_{C_{bc}}dze^{-L\left[\Lambda\left(n,z;\beta\right)+i\phi\left(n,z;\beta\right)\right]}\\
 & \approx & \frac{1}{\pi}\int_{1}^{1+\eta}dze^{-L\Lambda}\sin\left(L\phi\right),
\end{eqnarray*}
where as above $\eta\ll1$. Following similar steps to those done
in section B.1, one can find
\begin{eqnarray}
\Lambda\left(n,1+\delta z;\beta\right) & \approx & \Lambda\left(n,1;\beta\right)+b_{\Lambda}\delta z,\label{eq:APPB_Lam_expansion-1}\\
\phi\left(n,1+\delta z;\beta\right) & \approx & b_{\phi}\delta z^{\beta C-1}.\label{eq:APPB_Phi_expansion-1}
\end{eqnarray}
with
\begin{eqnarray*}
b_{\Lambda} & = & 1-\frac{n\zeta_{\beta c-1}}{\zeta_{\beta c}},\\
b_{\phi} & = & n\frac{\pi}{\Gamma\left(\beta c\right)\zeta_{\beta c}}.
\end{eqnarray*}
As $n<n_{c}=\zeta_{\beta c}/\zeta_{\beta c-1}$ this implies $b_{\Lambda}>0$
as expected. Following the same steps as in appendix B this implies
\begin{eqnarray}
Z_{N}(L,Ln;\beta) & \approx & \frac{e^{-L\Lambda\left(n,1;\beta\right)}}{\pi}\int_{0}^{\eta}d\delta ze^{-Lb_{\lambda}\delta z}\sin\left(Lb_{\phi}\delta z^{\beta C-1}\right)\nonumber \\
 & \approx & \frac{b_{\phi}\Gamma\left(\beta C\right)}{\pi L^{\beta C-1}b_{\Lambda}^{\beta C}}e^{-Lf_{n}\left(n,1;\beta\right)}.\label{eq:APPC_Z_N_branchcut_res}
\end{eqnarray}

\section{Numerical procedure for evaluating the partition function}

Here we explain the numerical procedure that is used to evaluate $Z_{M}\left(L,M;\beta\right)$
exactly. The partition function $Z_{M}$, with the boundary conditions $\sigma_1=1$ and $\sigma_L=-1$,
which we denote by $(+-)$, has the form
\begin{eqnarray*}
\fl Z_{M}^{(+-)}\left(L,M;\beta\right) & = & \sum_{\nu=1}^{\infty}\sum_{l_{1}=1}^{\infty}...\sum_{l_{N}=1}^{\infty}e^{-\beta\Delta\left(2\nu-1\right)}\left(\prod_{a=1}^{2\nu}\frac{1}{l_{a}^{\beta C}}\right)I\left(L=\sum_{a=1}^{2\nu}l_{a}\right)I\left(M=-\sum_{a=1}^{2\nu}\left(-1\right)^{a}l_{a}\right)\\
\fl  & = & \sum_{l_{1}=1}^{\infty}\frac{e^{-\beta\Delta}}{l_{1}^{\beta C}}\sum_{\nu=1}^{\infty}\sum_{l_{2}=1}^{\infty}...\sum_{l_{N}=1}^{\infty}e^{-\beta\Delta\left(2\nu-2\right)}\left(\prod_{a=2}^{2\nu-1}\frac{1}{l_{a}^{\beta C}}\right)\\
\fl  &  & \qquad\times I\left(L=l_{1}+\sum_{a=2}^{2\nu-1}l_{a}\right)I\left(M=l_{1}-\sum_{a=2}^{2\nu-1}\left(-1\right)^{a}l_{a}\right)\\
\fl  & = & \sum_{l_{1}=1}^{\infty}\frac{e^{-\beta\Delta}}{l_{1}^{\beta C}}Z_{M}^{(--)}\left(L-l_{1},M-l_{1};\beta\right),
\end{eqnarray*}
where $Z_{M}^{(--)}$ is the partition function corresponding to boundary
conditions $s_{1}=s_{L}=-1$ (so that the number of domains is odd).
A similar analysis for $Z_{M}^{(--)}$ reads
\begin{eqnarray*}
\fl Z_{M}^{(--)}\left(L,M;\beta\right) & = & \sum_{\nu=1}^{\infty}\sum_{l_{1}=1}^{\infty}...\sum_{l_{N}=1}^{\infty}e^{-\beta\Delta\left(2\nu-2\right)}\left(\prod_{a=1}^{2\nu-1}\frac{1}{l_{a}^{\beta C}}\right)I\left(L=\sum_{a=1}^{2\nu-1}l_{a}\right)I\left(M=\sum_{a=1}^{2\nu-1}\left(-1\right)^{a}l_{a}\right)\\
\fl  & = & \sum_{l_{1}=1}^{\infty}\frac{e^{-\beta\Delta}}{l_{1}^{\beta C}}\sum_{\nu=1}^{\infty}\sum_{l_{2}=1}^{\infty}...\sum_{l_{N}=1}^{\infty}e^{-\beta\Delta\left(2\nu-3\right)}\left(\prod_{a=2}^{2\nu-2}\frac{1}{l_{a}^{\beta C}}\right)\\
\fl  &  & \qquad\times I\left(L=l_{1}+\sum_{a=2}^{2\nu-2}l_{a}\right)I\left(M=-l_{1}+\sum_{a=2}^{2\nu-2}\left(-1\right)^{a}l_{a}\right)\\
\fl  & = & \sum_{l_{1}=1}^{\infty}\frac{e^{-\beta\Delta}}{l_{1}^{\beta C}}Z_{M}^{(+-)}\left(L-l_{1},M+l_{1};\beta\right).
\end{eqnarray*}
Hence we deduce the following coupled recursion relations
\begin{eqnarray}
Z_{M}^{(+-)}\left(L,M;\beta\right) & = & \sum_{l=1}^{\frac{L+M}{2}}\frac{e^{-\beta\Delta}}{l^{\beta C}}Z_{M}^{(--)}\left(L-l,M-l;\beta\right),\label{eq:APPC_Zpm_rec}\\
Z_{M}^{(--)}\left(L,M;\beta\right) & = & \sum_{l=1}^{\frac{L-M}{2}}\frac{e^{-\beta\Delta}}{l^{\beta C}}Z_{M}^{(+-)}\left(L-l,M+l;\beta\right).\label{eq:APPC_Zpp_rec}
\end{eqnarray}
The total chain size $L$ reduces in each step of applying these relations,
and hence convergence is guaranteed. The base of the recursion is
\begin{eqnarray}
\forall L:\quad Z_{M}^{(+-)}\left(L,\pm\left(L-2\right);\beta\right) & = & \frac{e^{-\beta\Delta}}{\left(L-1\right)^{\beta C}},\label{eq:APPC_Zpm_ini}\\
\forall L:\quad Z_{M}^{(--)}\left(L,-L;\beta\right) & = & \frac{1}{L^{\beta C}}.\label{eq:APPC_Zpp_ini}
\end{eqnarray}

\end{appendices}

\bibliographystyle{iopart-num}
%\bibliography{symm_long}
\providecommand{\newblock}{}

\end{document}